\documentclass[10pt, 3p, twocolumn]{elsarticle}

\usepackage{amsmath} %maths
\usepackage{array}
\usepackage{framed}
\usepackage{graphicx}
\usepackage{hyperref}

\graphicspath{{figures/}}

\usepackage{breqn}   % <--- keep in final version;  breqn is necessary for two-column format because of many long equations
\journal{Nuclear Instruments and Methods A}

\begin{document}

\begin{frontmatter}

\title{ Analytical modeling of pulse-pileup distortion using the true pulse shape; applications to \textit{Fermi-GBM} \tnoteref{t1} } 

\address[lab1] {CSPAR, University of Alabama in Huntsville.  301 Sparkman Drive, Huntsville, AL 35899 USA}
\address[lab2] {Department of Physics, University of Alabama in Huntsville.  301 Sparkman Drive, Huntsville, AL 35899 USA}

\author[lab1]{Vandiver Chaplin\corref{cor1}}
\cortext[cor1]{Corresponding author.  Electronic mail: chapliv@uah.edu.  Phone: 1-256-961-7514 }
\author[lab1]{Narayana Bhat}
\author[lab1,lab2]{Michael S. Briggs}
\author[lab1,lab2]{Valerie Connaughton}

\date{\today}

\begin{abstract}
Pulse-pileup affects most photon counting systems and occurs when photon detections occur faster than the detector's shaping and recovery time.  At high input rates, shaped pulses interfere and the source spectrum, as well as intensity information, get distorted.  For instruments using bipolar pulse shaping there are two aspects to consider: `peak' and `tail'  pileup effects, which raise and lower the measured energy, respectively.  Peak effects have been extensively modeled in the past.  Tail effects have garnered less attention due to increased complexity.  We leverage previous work to derive an accurate, semi-analytical prediction for peak and tail pileup including high order effects.  We use the pulse shape of the detectors of the \textit{Fermi Gamma-ray Burst Monitor}.  The measured spectrum is calculated by expressing exposure time with a state-space expansion of overlapping pileup states and is valid up to very high rates.  The model correctly predicts deadtime and pileup losses, and  energy-dependent losses due to tail subtraction (sub-threshold) effects.  We discuss total losses in terms of the true rate of photon detections versus the recorded count rate.
\end{abstract}


\tnotetext[t1]{   \url{http://dx.doi.org/10.1016/j.nima.2013.03.067}  }

\end{frontmatter}

\section{Introduction}

%\begin{linenumbers}

Pulse pileup affects X- and gamma-ray counting systems in the presence of high-intensity sources.  At high rates the \emph{detected} events become increasingly different from those \emph{registered} by the instrument.\footnote{Terms such as \textit{detected, input} will refer to the photons interacting in the detector volume.  Terms such as \textit{registered, recorded, measured}, \textit{observed} will refer to the subset of photons and pileup events which are finally counted by the instrument}  Detected events produce shaped electronic pulses that require some minimum time interval to measure the pulse height, and additional time to recover.  The total time elapsed is the `pulse window'.  Additional detections in this interval will modify the expected pulse shape and distort the inferred pulse-height information and detection rate.  In this paper we elaborate on previous methods to develop a correction for pileup using the true bipolar pulse shape, while also requiring fewer assumptions \cite{taguchi:2010}.  We apply this method to the space-borne gamma-ray instrument \textit{Fermi Gamma-ray Burst Monitor} (GBM).  GBM consists of a set of inorganic scintillation detectors and is one of two instruments on the \textit{Fermi} gamma-ray space telescope.  An overview of the GBM detector hardware, science mission, and the context for pulse-pileup studies is described in section \ref{Fermi-GBM}.

In bipolar-pulsed instruments pileup has two effects on the measured pulse spectrum, depending on the arrangement of pileup events.  \emph{Peak} effects arise from the addition of \emph{positive} pulse sections, causing a single apparent high-energy count.  \emph{Tail} effects occur when events are detected in the negative tail of a previous event, causing a second count with reduced apparent energy.  For pulse shapes with large negative swings, modeling the tail pileup is especially important since losses (non-registered events) will occur if the summed peak is below threshold.  Such losses depend on the input spectrum, a dependence which gets stronger as the bipolar peak-to-peak ratio approaches unity.

With some assumptions the distorted spectrum can be predicted in terms of the true input rate and spectrum.  Cano-Ott \cite{CanoOtt1999488} and Danon \cite{Danon2004287} model peak effects in the context of unipolar pulses.  Taguchi \cite{taguchi:2010} model both the peak and tail effects in an X-ray counter with a bipolar pulse.  The approach generally taken is to model the ``pileup response'' of the detector, which is a conditional probability (a.k.a. ``likelihood function'') of recording a certain energy $\varepsilon$ resulting from pileup of two events, with energies $E_0$ and $E_1$:  
\begin{equation*}
\text{Pr}(\varepsilon | E_0, E_1 )
\end{equation*}
Calculating such functions requires basic information about the hardware response of the detector to photon energy deposition, and a statistical model for the time-separation between detections.

Assuming linearity in photo-current production and pulse shaping, the detector's observed analog signal is prescribed using a superposition of individual pulses.  Then the likelihood functions are calculated assuming Poisson distributed events.  The pileup spectrum is written using ``total probability'', an integral over measured energies (or sum over channels) of the pileup response times the probability of input energies (i.e., the input spectrum).  Sections \ref{PeakEffect} and \ref{TailEffect} derive the likelihood functions we use for modeling GBM pileup.

A common simplification is to forego the true pulse shape when calculating the pileup-energy likelihood, and instead use a mathematical approximation.  This has the advantage that likelihoods can have  closed form expressions.  For example, \cite{taguchi:2010} use a triangular or `delta' approximation, such that the summed peak-time (and amplitude) is given by simple geometry, and its dependence on the event separation is algebraically invertible.  In section \ref{1stOrderPeakSpecDerive} we demonstrate why invertibility is necessary for a closed form likelihood expression.  The disadvantage in approximating is that the true pulse shape gives significantly more accurate predictions than the approximate shapes. \cite{CanoOtt1999488} demonstrate this by evaluating their model with an accurate Gaussian pulse and then with several approximate shapes.  They compare results with Monte Carlo simulation and show that only the true pulse shape is accurate over the entire energy range.  They include only first order peak effects and reach a maximum pileup fraction well below what is expected in our applications.

The technique of \cite{taguchi:2010} allows for modeling high order pileup effects using an iterative approximation.  Tail modeling uses a convolution method rather than the pileup-likelihood technique used for the peak effect.  This admits a number of simplifying assumptions about the timing and energy distribution of tail events and avoids complications in calculating additional likelihood functions.  However, when applied to the true pulse shape and specific input spectral shapes in our application, we found such assumptions did not yield sufficiently accurate results.  This is perhaps due to a larger negative amplitude in our instrument, making the distorted spectrum more affected by tail pileup.

The peak pileup method described in this paper is largely an application of the method of \cite{taguchi:2010} to the GBM pulse shape.  However, we present a different method for tail modeling which extends the likelihood function treatment to tail regions, allowing us to model the spectral and timing randomness of tail events.  This method also accounts for events occurring in a fixed deadtime that is separate from the peak pileup interval. 

A novelty of this approach is that the total spectrum is obtained by partitioning time into overlapping pulse windows.  Pulse windows are written as states of a Poisson process, each one having an associated peak and tail spectrum.  In section \ref{sec:PPU} a state is defined by the number of events in the window, with pileup states having a non-zero number (the \emph{pileup order}).  In sections \ref{PeakEffect} and \ref{TailEffect}, the recorded energy distributions are derived based on the number of events per state and how they are arranged into three sub-intervals of the pulse window.  The total time in which there is a non-zero analog signal is then the union of all pulse windows, and the total spectrum is the corresponding union of peak and tail spectra per state.  This is expressed as superposition of pulse states minus overlapping terms (section \ref{FullExpansion}).  We compare the model to Monte Carlo simulations in section \ref{MonteCarloSection}.

As a final motivation for using the true bipolar pulse shape, we note there is only a small increase in computational overhead.  Despite the apparent convenience of closed form likelihood functions, the actual difference in utility between them and a numerical function is small.  Most spectra are sufficiently complicated that the total probability of likelihood times spectrum over energy is not analytically integrable, thus numerical evaluation is required.  This is especially the case for non-ideal detectors, which have a complicated instrument response $R(E, E_{\gamma})$, and the deposited energy spectra are already expressed numerically.  Other than some storage and memory access overhead, computation of the pileup spectra requires about the same number of operations in either approach.  As will be shown, for the constant Poisson process the likelihood functions are independent of source intensity and thus can be calculated \emph{a priori}, stored in memory, and read by a separate program as required.  In section \ref{Evaluation} we describe a simple numerical calculation of the likelihood functions.

%%%%%%%
%
%   SECTION
%	GBM info
%
%%%%%%%
\section{Fermi-GBM} \label{Fermi-GBM}

GBM is a gamma-ray counting instrument aboard the \textit{Fermi Gamma-ray Space Telescope}.  It was launched in June 2008 and has been in near-continuous operation since the start of normal operations one month later.  GBM consists of 12 sodium-iodide (NaI) and two bismuth-germanate (BGO) detectors producing time and energy resolved data sets.  NaI detectors have an effective energy range from approximately 8 keV to 1 MeV, and BGOs from 200 keV to 40 MeV.  Pulse heights are digitized into 128 pseudo-logarithmic spectral channels per detector.  Detector gains are calibrated such that channel boundaries lie between 0 to 5 volts.  More information on GBM hardware, detectors, and electronics is available in \cite{Meegan}.   Instrument calibration (the relationship between peak voltage and input energy) studies are described in \cite{Bissaldi2009}. \footnote{For simplicity the figures in this paper are calculated using BGO channel energies that are approximated to be logarithmic: $E_m = (150 \text{keV}) (1.04557)^m$ gives the start of the $m^{th}$ energy channel in keV. Plots can be approximately ``converted'' to the NaI energy range for a channel $m$, using $E_m = (4 \text{keV})*(1.04408)^m$.  Actual channel energies are not precisely logarithmic.}

Each detector has its own shaping and digitization firmware for performing pulse height analysis (PHA) of detected events.  Pulse-pileup, when it occurs, is on a detector-by-detector basis rather than an integrated signal.  The pulse shape has a finite width that requires $~0.8 \mu s$ to register a single event and an additional $~3.5-4 \mu s$ for baseline recovery (section \ref{GBMPulseShape}).  Generally speaking, pulse pileup occurs when the separation between detected events is smaller than the registration + recovery time.   The negative peak amplitude is about 70\% of the positive peak amplitude.  To help mitigate pileup effects, a fixed deadtime of $\tau = 2.6 \ \mu s$ is applied once a peak has been found, preventing pulse measurement during most of the recovery.  Pulse pileup however extends the time for baseline recovery and $\tau$ becomes insufficient, adding tail distortions to the measured spectrum.

GBM is primarily an astrophysics experiment designed to detect transient cosmological sources such as gamma-ray bursts (GRBs).    Detectors are uncollimated and typical source + background rates are $1-2 \times 10^3$ counts per second (cps) per detector.  In the majority of cases a simple non-paralyzable deadtime correction is sufficient, but pileup is potentially problematic for several types of sources observed.  Depending on the application, pileup effects would become non-negligible at an \emph{input} rate ($\lambda_0$) around $10^5$ cps in one detector, corresponding to an \textit{observed} counting rate ($\lambda_{Rec}$) around 60,000 - 70,000 cps in that detector.  %The GBM Data Processing Unit (DPU) was designed to record at a maximum rate ($\lambda_{Rec}$) of 100,000 cps and a maximum of 600,000 cps from all  14 detectors together.

In NaI detectors soft-gamma repeaters (SGRs) can have peak recorded rates near this maximum \cite{AJvdh_2012}.  Solar flares are also routinely observed at these levels.  Terrestrial gamma-ray flashes (TGFs) typically peak around $\lambda_{Rec} \sim 10^5$ cps in BGO detectors \cite{Briggs2010}.  In such cases the true intensity is much greater than what is inferred from a simple deadtime correction.  At these levels a substantial fraction ($\sim$65\% - 75\%) of input photons pile-up (section \ref{PileupLosses}), and the subset of resulting counts is highly distorted.  Modeling the spectrum and energetics of such high-intensity transients thus requires understanding pileup distortion in GBM, and our goal is to have a fast, flexible numerical prediction which might be employed in spectral analysis.  

The model we present gives accurate results up to very high values of the input rate $\lambda_0$.  Demonstrations of spectral distortion versus rate are given in section \ref{Evaluation}.  The relation between $\lambda_{Rec}$ and $\lambda_{0}$ is considered in more detail in section \ref{PileupLosses}.

%%%%%%%
%
%   SECTION
%	Pulse shape
%
%%%%%%%
\section{Pulse shape}\label{GBMPulseShape}

GBM has well-studied electronic and detector properties that allow us to model its behavior with mathematical expressions. Shaping circuits make pulses whose first zero-crossing is fixed relative to the pulse start, regardless of the total photocurrent.  We assume the zeroth order shape can be written in a separable form where the shape $f(t)$ is modulated by a scalar that depends on the amount of input energy.  Thus a given pulse can be written $V(t) = v_0 f(t)$, where $v_0$ is an energy dependent coefficient, and $f(t)$ is the basic pulse shape.  Using data collected in pre-launch testing, we fit the function below to sampled analog signal forms, for a range of input energies:
\begin{equation}
f(t) =  K (c_1 * t^{\alpha} - c_2 * t^{\beta} ) e^{-\gamma t}
\label{eq:PulseShape}
\end{equation}
where $c_1, c_2, \alpha, \beta, \gamma$ are fitted constants that parameterize the pulse shape.  $K$ is a normalizing constant such that $f(t)$ is unity at its peak (section \ref{ZerothOrderPeakTime}).

The best-fit shape parameters are $c_1 = 26.0, c_2 = 31.0, \alpha=1.27, \beta=3.5, \gamma = 2.6, K = 0.41023$, for the time $t$ in microseconds.  The figures and specific results in this paper refer to equation \eqref{eq:PulseShape} with these parameters.  However, as evidence for the generality of the technique, we have tried a wide range of parameter values and find the model, when compared to Monte Carlo simulations using the same pulse shape, to be equally accurate in all cases.

%%%%%%%%%
%  Partition of the pulse shape
%%%%%%%%%
%
% Figure: pulse partition 1
%
\begin{figure} [ht!]
\centering
\includegraphics[width=\linewidth]{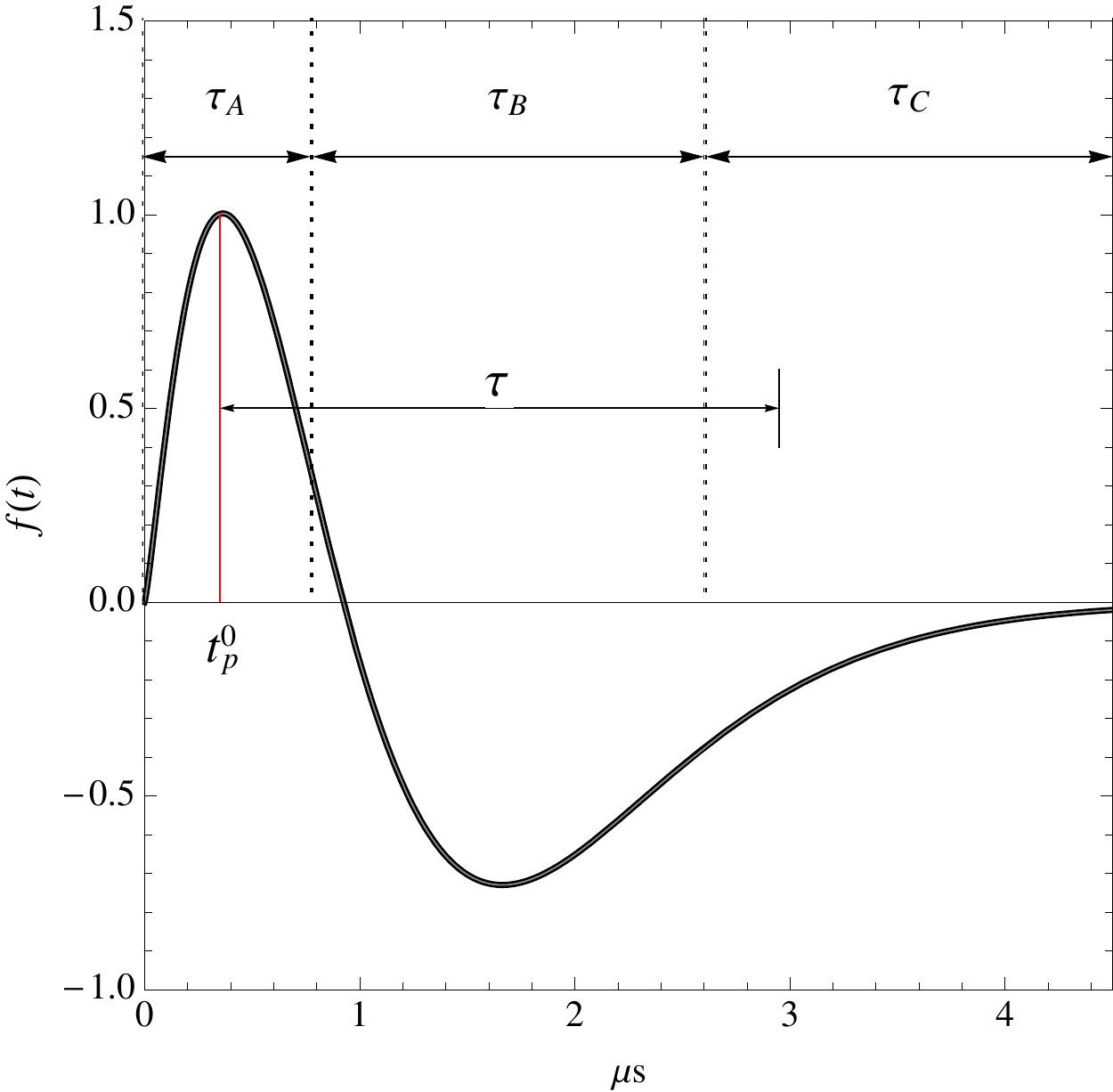}
\caption{A single unit pulse showing the zeroth-order peak time $t^0_p$, the applied deadtime $\tau$, and the partitions A, B, and C.  The pulse ``window'' is $\Delta\tau = \tau_A + \tau_B + \tau_C =$ 4.5 $\mu s$.}
\label{fig:UnitPulseIntervals}
\end{figure}

\subsection{Partition of the pulse window} \label{PulsePartitionSection}
\begin{figure*}[t]
\centering
\includegraphics[width=\linewidth]{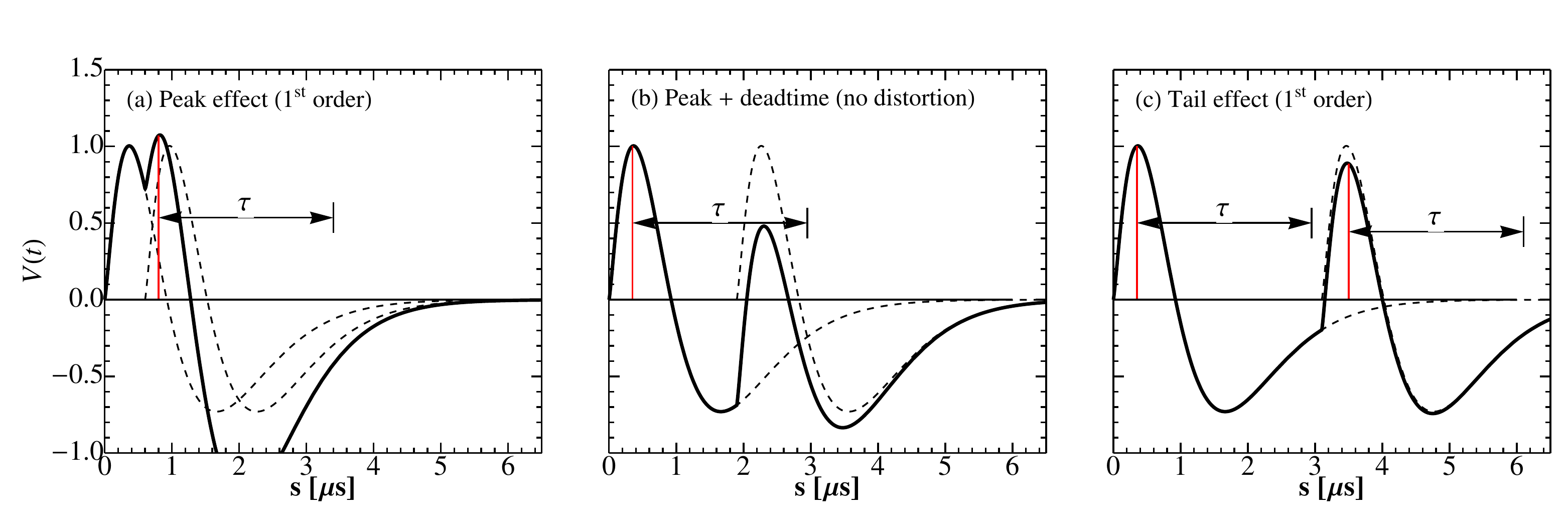}
\caption{The three cases of `first order' pileup, $\langle 100 \rangle$, $\langle 010 \rangle$, and $\langle 001 \rangle$, showing the measured peak for two events of equal energy, and the dead time $\tau$ as imposed by GBM hardware.  (a) shows the \emph{peak} effect, and (c) the \emph{tail} effect.  Panel (b) depicts a nominal case where one count is accurately measured and the next is lost.  Typically this is not regarded as `pulse pileup' as there is no associated spectral distortion of the pulse height, only the count rate, which can be corrected by simpler means.   }
\label{fig:ThreePanel1}
\end{figure*}

The pulse window initiated by detection of an initial event will be partitioned into three regions, each corresponding to different distortion effects.  We label them `A', `B', and `C', having widths $\tau_A, \tau_B$, and $\tau_C$, respectively (figure \ref{fig:UnitPulseIntervals}).  A and C will also be called the \emph{peak} and \emph{tail} regions.  In our model, the particular distortion observed will be parameterized in terms of the number of events in each sub-interval, and pileup defined as the presence of additional photons in any of these sub-intervals.  Choice of widths depends on the specifics of the pulse shape, the algorithm which measures pulse height, and the deadtime implementation.  However, we can give the following generic definitions for each region, based on the case of first-order pileup.  Figure \ref{fig:ThreePanel1} depicts the three cases of first order pileup, described below.

Suppose two photons are detected within a pulse-width of each other.  The initial pulse we call the `zeroth' event, and let it begin at time $t_0 = 0$.  Then the next event, beginning at $t_1$ and having an associated voltage $v_1$, can occur in one of the three intervals A, B, or C.  If $t_1$ is in A, peaks will add and a `summed' pulse height appears which is a function of $v_0,v_1$, and $t_1-t_0$.  If $t_1$ is in B, $v_0$ is measured correctly, and the second peak is shifted down due to the negative baseline.  In the case of GBM this coincides with deadtime, so there is no second count.  Finally if $t_1$ is in C, there is a second measurement, but the peak is shifted down from its nominal value $v_1$.  Thus the widths are defined as follows.  `$\tau_A$' is the minimum time required to resolve the pulse height of one event.  For our model to apply it should be (approximately) $\leq$ the positive signal width.  `$\tau_B$' is a dead interval: if the second event occurs in this region it is not measured, but will influence the region C.  `$\tau_C$' is a live region, but lasts until the zeroth pulse is baseline-recovered.  

In the specific context of GBM, the width $\tau_A$ of the first region is the time-to-max $t^0_p$ plus a buffer following the peak: 
\begin{equation*}
\tau_A = t^0_p + \tau_{\text{Buff}} = 0.78\ \mu s
\end{equation*}
This buffer is part of the instrument's digital peak finding algorithm, which requires a decreasing voltage for a sequence of digital samples before registering a pulse height.\footnote{This scheme is designed to eliminate measurements due to electrical noise, and to apply an energy-independent dead time for each event.  GBM samples analog pulses with a period of 0.104 $\mu s$. The buffer is programmed to be four samples long, so $\tau_{\text{Buff}} = 0.416 \mu s$.  The deadtime results from waiting an additional 21 samples after a peak is found and buffered, so $\tau = (4+21)*0.104 = 2.6 \ \mu s$. }  If additional events occur before the buffer expires (i.e., anywhere in region A), they will cause peak-pileup. \cite{Theis2006}

Events in part B are lost.  Applied deadtime $\tau$ contains the B interval.  $\tau_B$ smaller than $\tau$ because a pulse occurring on its periphery might carry into the live time and be measured.  Likewise, a pulse occurring in the post-peak buffer would be measured. Thus $\tau_B$ is the deadtime $\tau$ minus this buffer on the left, and the time-to-max on the right:
\begin{equation*}
\tau_B =  \tau - (t^0_p + \tau_{\text{Buff}}) =  \tau - \tau_A = 1.82\ \mu s
\end{equation*}
where $\tau = 2.6\ \mu s$ is the GBM deadtime.

Region C is cutoff at an arbitrary point where the baseline (due to the initial pulse) can be considered to be recovered.  But the instrument is also live here, and peak pileup from additional events is possible.  Thus we \emph{also} require it to be short enough that only a single peak can be recorded in this region.  For GBM modeling we use the value $\tau_C = 1.9\ \mu s$, to give a total pulse window of width
\begin{equation}
\tau_A + \tau_B + \tau_C = \Delta \tau = 4.5 \mu s
\end{equation}

Predictions from the model have some dependence on the value of $\tau_C$.  We examined this by trying both larger (up to 6 $\mu s$) and smaller (down to 4 $\mu s$) pulse window sizes, varying $\tau_C$ only.  The model's accuracy was not appreciably different for window sizes between $4.5 - 5.0 \mu s$.  For smaller window size, the amount of tail distortion becomes under-predicted when compared to MC simulations (i.e., not enough low-energy counts are predicted).  For larger sizes, it becomes over-predicted.  This is not surprising, since the probability of tail distortion increases with increasing $\tau_C$ (see equation \eqref{eq:StateProb}). For a successful model $\tau_C$ must be a good approximation to the instrument's actual recovery time, such that the computed probability of tail distortions is equal (within statistics) to the fraction of tail counts from simulation.  The value used in this paper was not optimized for model accuracy but chosen \textit{a priori} based on the known pulse shape and the constraints given in the previous paragraph.

% FIGURE 2
% Three-panel PPU1
%

% FIGURE 3
% Three-panel PPU1
%
\begin{figure*}[ht]
\centering
\includegraphics[width=\linewidth]{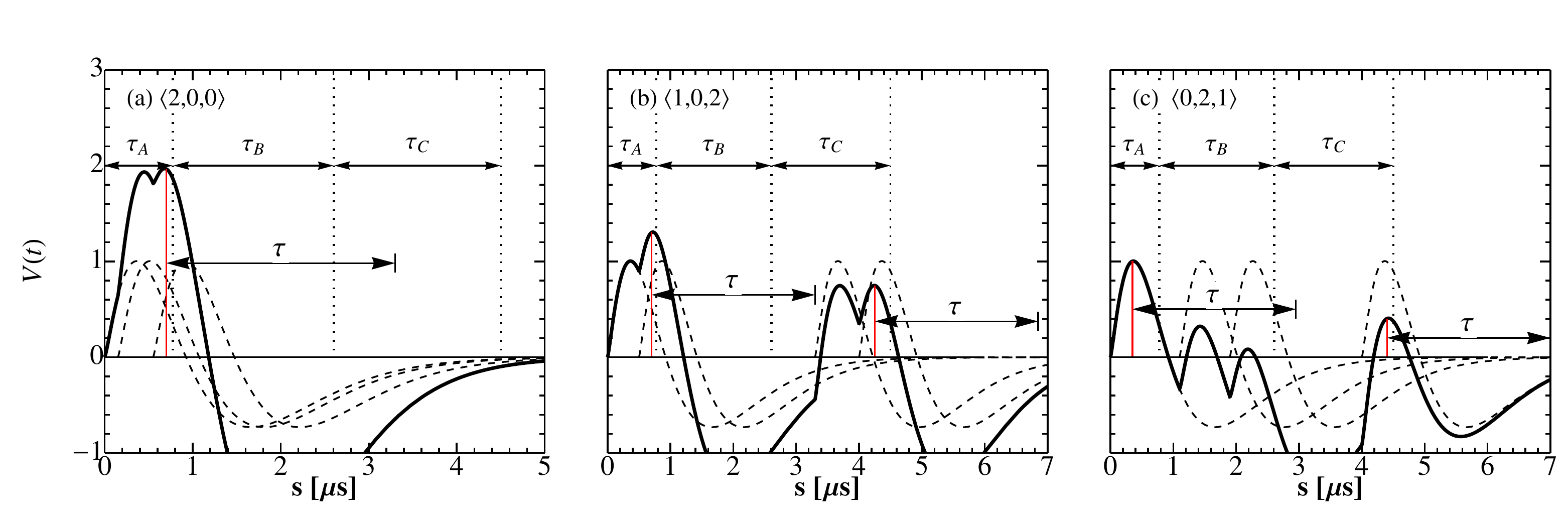}
\caption{ Higher order pileup examples, with the A-B-C partitions shown. (a) second order peak pileup.  (b) third order pileup, with peak \emph{and} and tail effects.  (c) a third order case of the deadtime+tail effect.  Recorded pulse height in C depends on the tail from pulses in A and B. }
\label{fig:ThreePanel2}
\end{figure*}

%%%%%%%
%
%   SECTION
%	PPU definitions
%
%%%%%%%
\section{Pulse-pileup}\label{sec:PPU}

Having partitioned the pulse shape into three adjacent regions, we can now describe pulse-pileup in terms of the number of events in each one.  Assuming a zeroth event to begin the window, one or more added events in region A constitutes \emph{peak pileup}.  One or more events occurring in region C cause \emph{tail pileup}, or the tail effect.  Additionally, pulses in B contribute to baseline reductions for the C events, and this is a more extreme category of tail effects.  Together these produce artificial raising and lowering distortions in the measured spectrum, which it is our goal to model.

Assuming a zeroth event with the A-B-C partitions, the \emph{state} or \emph{configuration} of pulse-pileup can be represented using three non-negative integers $a, b, c$ which give the number of additional events in each interval (excluding event 0).  We define the symbol $\langle a,b,c \rangle$ to describe the pileup state in the pulse window $\Delta \tau$.  The total \emph{order} of pileup is the number of additional events, such that order zero is a single count without pileup, first order is one added event in A, B, or C, etc.  Thus the generic configuration $\langle a,b,c \rangle$ represents the case of an incident event (the zeroth event) followed by $(a+b+c)$ events within the window.  In later sections we address the probability and independence of pileup states, but for now we need only the definition:

%%%
% EQUATION:  State definition
%
\begin{align*}\label{eq:StateDefinition}
&\text{Event } \langle a,b,c \rangle\ \\
&\ \ \equiv\ \text{$a$ pulses in A, $b$ in B, and $c$ in C}
\end{align*}
In general this represents the event of $(a+b+c)^{th}$ order pileup, where $(a+b+c) \geq 0$, and $\langle 0,0,0 \rangle$ is the event of a single recorded count, accurately measured.  In this scheme the maximum number of measured counts is two (one from the peak, one from the tail).

Figure \ref{fig:ThreePanel1} demonstrates the simplest case: \emph{first-order} pileup.   Figure \ref{fig:ThreePanel2} gives several examples of higher order pileup, with the pulse partitions shown.  In the following sections we calculate pulse height distributions corresponding to each state $\langle a,b,c \rangle$, and an input energy spectrum.  The peak spectrum will depend only on $a$, but the tail spectrum depends on all three orders.

One consequence of pulse pileup is that the deadtime is extended beyond its the nominal value $\tau$.  For example, pileup in region A results in a summed peak at a time $\geq$ the single-pulse peak time $t^0_p$.  And since the deadtime is imposed from this registered peak, it effectively migrates forward, contributing to \emph{paralyzable deadtime} \cite{WRLeo}.  Strictly speaking the pulse partitions should have a corresponding shift.  Instead we use a semi-paralyzable model where the summed peak time (section \ref{GBMPeakTime}) can wander into region B, but the partitions and deadtime remain fixed.  Figures \ref{fig:ThreePanel2}(a-c) show how the registered peak shifts forward while ABC partitions are imposed from the zeroth event. The result is shown to be accurate when compared to Monte Carlo simulations.

%%%%%%%
%
%   SECTION
%	Interval density
%
%%%%%%%
\section{Interval Distribution}
Within fixed intervals the interval distribution (i.e., the probability density (PDF) of separation time between events) can be calculated under the assumption of a fixed-rate Poisson process.  This derivation is well-known, appearing in \cite{taguchi:2010} among other places, but because of its centrality in the calculation of pileup likelihood we also present it.  For the Poisson process of rate $\lambda$, the PDF of the separation $s$ between any event and the next is 
\begin{equation}
f_s(s) = \lambda e^{-\lambda s}
\end{equation}
which is defined on $s \in (0, \infty)$.  Define the separation of the $i^{th}$ event from the $(i-1)^{th}$ as 
\begin{equation}
s_i \equiv t_i - t_{i-1}
\end{equation}
where $t_i$ is the time of event $i$.  Now suppose a finite interval ranging from 0 to $\Delta t$, with the zeroth event outside the interval at $t_0 = 0$.  Then the probability of \emph{exactly one} event \emph{in} the interval is given by the (joint) probability that $s_1 \leq \Delta t$ \textbf{and} $ s_1 + s_2 > \Delta t$, and the second condition is equivalent to $s_2 > \Delta t - s_1$.

\begin{dgroup*}
\begin{dmath*}
\text{Pr}( s_1 \leq \Delta t \text{ and } ( s_2 > \Delta t - s_1)\ ) = \int \limits_0^{s_1} \lambda e^{-\lambda s_1} ds_1 \int \limits_{\Delta t-s_1}^{\infty} \lambda e^{-\lambda s_2} ds_2
\end{dmath*}
\begin{dmath}
=\ \lambda s_1 e^{-\lambda \Delta t} 
\end{dmath}
\end{dgroup*}
Differentiating in $s_1$ gives the joint probability density $f_s( s_1\  ,1 )$, whose general form is $f_s( s_1\  ,n )$ for the random variables $s_1$ and $n$ (number of events in $\Delta t$)
\begin{equation}
f_s( s_1 , 1) = \frac{d}{ds_1}\text{Pr} =  \lambda e^{-\lambda \Delta t}
\end{equation}
The \emph{conditional} probability density, $f_{s|n}(s)$, is given by dividing the joint density $f_s(s_1, 1)$ by the probability that $n=1$, a procedure equivalent to normalizing over the interval $[0,\Delta t]$:

\begin{dgroup*}
\begin{dmath}
f_{s|n=1}(s)= \frac{ f_s(s_1 , 1) }{ \int_0^{\Delta t} f_s(s_1^{\prime} , 1) ds_1^{\prime} }  = \frac{ \lambda e^{-\lambda \Delta t}}{  \lambda \Delta t e^{-\lambda \Delta t} } = \frac{1}{\Delta t}
\end{dmath}
\end{dgroup*}
which is a constant, and is independent of $\lambda$. \cite{Grimmet}

% FIGURE 4
%  2nd Order Interval Diagram
%

\begin{figure}[t]
\centering
\includegraphics[width=\linewidth]{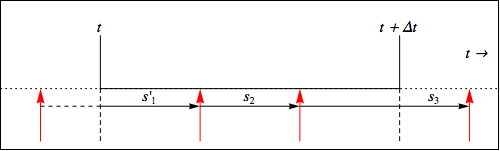}
\caption{$2^{nd}$ order pileup in the offset interval $[t, t+ \Delta t]$, showing events $t_0, t_1, t_2, t_3$.  If we assume finite pulses occur at each time, events $t_1,t_2$ are piled-up, with possible tail effects from $t_0$.  $s_1 = s_1^{\prime} + t - t_0$, and the random variables $s_1^{\prime}, s_2$ are identically distributed. }
\label{fig:SecondOrderInterval}
\end{figure}

For exactly two events in the interval, the condition is  $s_1+s_2 \leq \Delta t$ \textbf{and} $ s_1 + s_2 + s_3> \Delta t$ (e.g. figure \ref{fig:SecondOrderInterval}).  This can also be written  $s_2 \leq \Delta t - s_1 $ \text{and} $ s_3> \Delta t - (s_1 + s_2)$.
Then
\begin{dgroup*}
	\begin{dmath*}
		\text{Pr}( s_1\leq \Delta t\ \text{for 2 in $\Delta t$} ) = \int \limits_0^{s_1} \lambda e^{-\lambda s_1} ds_1 \\ \times \int \limits_0^{\Delta t - s_1}  \lambda e^{-\lambda s_2} ds_2 \int \limits_{\Delta t- (s_1+s_2) }^{\infty} \lambda e^{-\lambda s_3} ds_3
	\end{dmath*}
	\begin{dmath}\label{eq:2ndOrderJointDist}
		=\ \lambda^2 e^{-\lambda \Delta t} (s_1 \Delta t - \frac{s_1^2}{2})
	\end{dmath}
\end{dgroup*}

Assuming a definite state of $n=2$ in $\Delta t$ gives the second order conditional density:
\begin{dmath}\label{eq:2ndOrderConditionedDist}
	f_{s|2}(s_1)  = \frac{ f_s( s_1, 2) } { \int_0^{\Delta t} f_s(s_1^{\prime}, 2) ds_1^{\prime}} = \frac{2}{\Delta t^2}(\Delta t - s_1)
\end{dmath}
Because in this process event separations are independent, identically distributed random variables, $s_1, s_2$ follow the same probability density $f_{s|2}$.  
For the general case of $n$ events, the joint density function $f_s( s_1, n)$ is calculated as in equation \eqref{eq:2ndOrderJointDist} and the conditional density $f_{s | n}(s)$ just like \eqref{eq:2ndOrderConditionedDist}, which can be shown by induction to be, for all $n \geq 1$,
\begin{equation}\label{eq:nthOrderIntervalDist}
	f_{s | n}(s_1)  =  \frac{n}{\Delta t^{n}}(\Delta t - s_1)^{n-1}
\end{equation}

\subsection{ Offset Intervals } \label{MultipleIntervals}
The above density was derived assuming $t_0=0$, but equation \eqref{eq:nthOrderIntervalDist} turns out to have the same form regardless of the position of the previous event $t_0$ before the interval.  Consider the second-order integral in equation \eqref{eq:2ndOrderJointDist}, but now let the interval begin at some arbitrary time $t \geq 0$ and cover $t$ to $t + \Delta t$.  Let the zeroth event occur at an arbitrary time $t_0 \leq t$.  Now the $2^{nd}$ order pileup conditions, for $t_1,t_2$ in the interval, are:
\begin{align*}
t - t_0 < s_1 < (t+\Delta t) - t_0\\
s_1 + s_2 \leq (t+\Delta t) - t_0\\
s_1 + s_2 + s_3 > (t+\Delta t) - t_0
\end{align*}
An example is shown in figure \eqref{fig:SecondOrderInterval}.  Using the symbols $T_A \equiv t-t_0$ and $T_B \equiv  (t+\Delta t) - t_0$ for shorthand, the second-order joint probability is
\begin{dgroup*}
	\begin{dmath*}
		\text{Pr}( s_1, 2 ) = \int \limits_{T_A}^{s_1} \lambda e^{-\lambda s_1} ds_1 \\ \times \int \limits_0^{T_B - s_1}  \lambda e^{-\lambda s_2} ds_2 \int \limits_{T_B- (s_1+s_2) }^{\infty} \lambda e^{-\lambda s_3} ds_3
	\end{dmath*}
	\begin{dmath}
		=\ \frac{\lambda^2 e^{-\lambda T_B}}{2} \Bigl[ T_A^2 - s_1^2 + 2 T_B(s_1 -T_A) \Bigr]
	\end{dmath}
\end{dgroup*}
Differentiating and normalizing gives the conditional density:
\begin{dgroup*}
	\begin{dmath}
		f_{s|2}(s_1)  = \frac{  2 (T_B - s_1) } {( T_B- T_A)^2  } = 2\frac{   ( [t+\Delta t - t_0] - s_1 ) } {(\Delta t)^2  }
	\end{dmath}
	\begin{dmath*}
			 = \frac{  2 (\Delta t - [s_1 - (t - t_0)] ) } {(\Delta t)^2  }
	\end{dmath*}
\end{dgroup*}
Taking just the portion of $s_1$ that is in the normalization interval, $s_1^{\prime} = s_1 - (t-t_0)$, and substituting this into the the above, 
\begin{equation}
f_{s|2}(s_1^{\prime}) = \frac{  2 (\Delta t - s_1^{\prime} ) } {\Delta t^2  }
\end{equation}
where $0 < s_1^{\prime} \leq \Delta t$,  which is exactly the result when $t_0$ is at the start of the interval, equation \eqref{eq:2ndOrderConditionedDist}.  The subtracted term $(t-t_0)$ is a constant, so $s_1$ and $s_1^{\prime}$ have the same distribution except for this offset.  This generalizes to the result
%
%%%%
% EQUATION 
%		I^n   Nth Order Interval Dist
%%%
\begin{equation}\label{eq:nthOrderArbitraryIntervalDist}
	f_{s|n}(s)  =  \frac{n}{\Delta t ^{n}}(\Delta t  - s)^{n-1} 
\end{equation}
for $n$ events in an arbitrary interval $[ t,  t + \Delta t ] $, and $0 \leq s  \leq \Delta t$.  Using this result we can write the probability densities assuming a definite pileup state $\langle a,b,c \rangle$:
\begin{framed}
\begin{align}
f_{s|a}(s)  &=  \frac{a}{(\tau_A) ^{a}}(\tau_A  - s)^{a-1}  \label{eq:nthOrderPeakIntervalDist}  \\
f_{s|b}(s)  &=  \frac{b}{(\tau_B) ^{b}}(\tau_B  - s)^{b-1}\\
f_{s|c}(s)  &=  \frac{c}{(\tau_C) ^{c}}(\tau_C  - s)^{c-1} \label{TailIntervalDist}
\end{align}
\end{framed}

%%%%%%%%%%
% SECTION
%
%	PEAK EFFECT
%
%%%%%%%%%%%
\section{Peak pileup effect}\label{PeakEffect}
%In this section we apply the peak pileup technique of \cite{taguchi:2010}.  But where they use triangular pulses to simplify the math and find closed form equations, we provide the derivation in the context of the true pulse shape, equation \eqref{eq:PulseShape}.  Using the true pulse shape has the benefit of increasing the model accuracy, at the cost of some mathematical complexity.

The peak modeling technique is an application of the one presented in \cite{taguchi:2010}.  The method is to first derive expressions of the form $\text{Pr}(\varepsilon | E^{\prime}, E^{\prime\prime})$, which give the likelihood of recording energy $\varepsilon$ when two events are detected with energies $E^{\prime}, E^{\prime\prime}$, within a time $\tau_A$ of each other.  For first-order the model is accurate without any approximation.  At higher orders an iterative approximation is used since the dimensionality (i.e., number of random variables) becomes large.  Then the total probability of  $\varepsilon$ is given in terms of the pileup likelihood and the input spectrum.
% FIGURE 5
%  Peak pileup 1
%
\begin{figure}[!h]
\centering
\includegraphics[width=\linewidth]{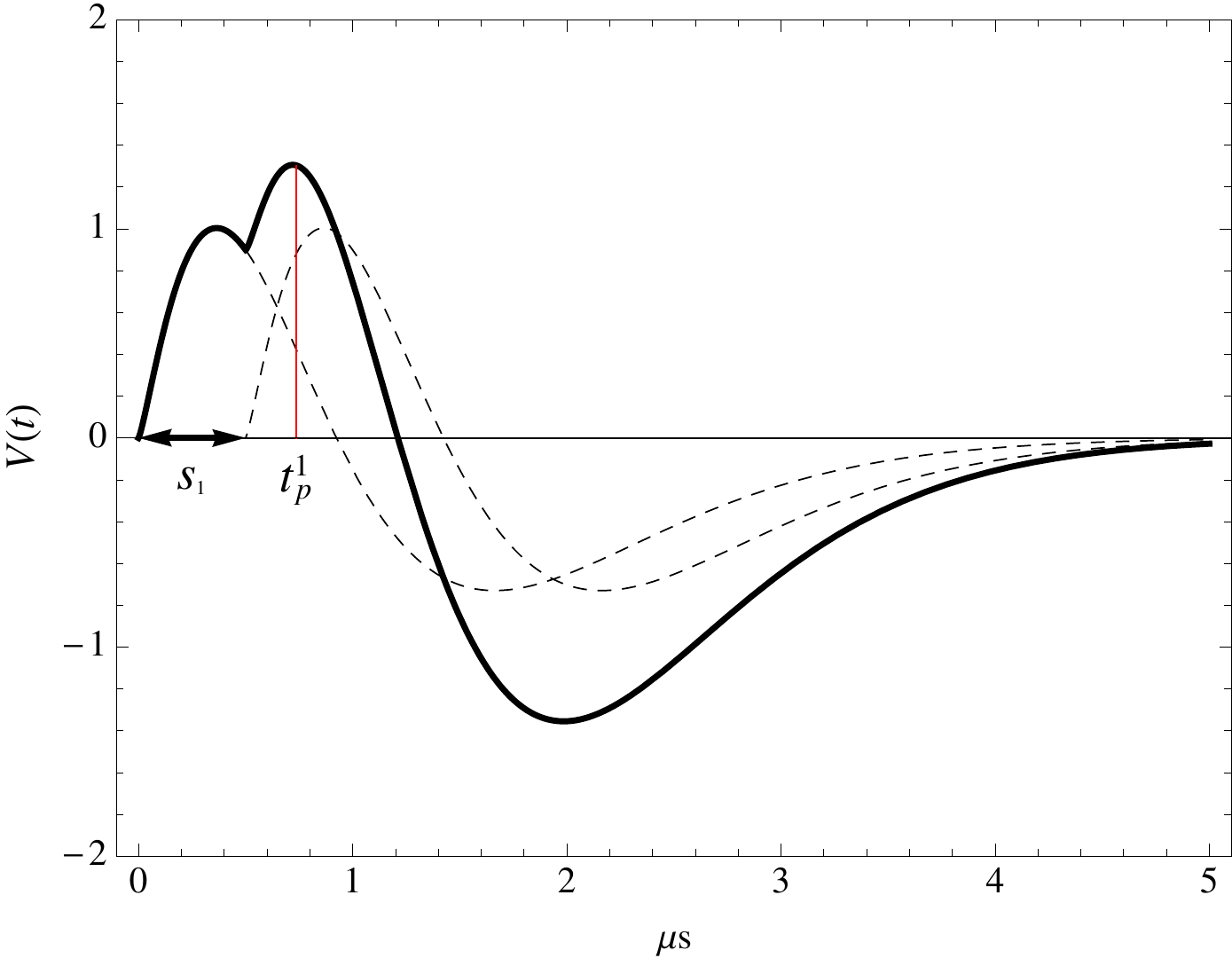}
\caption{A case of $1^{st}$ order peak pileup, showing the modeled peak at $t_p^1$ .}
\label{fig:SummedPeak1}
\end{figure}

%
% ZEROTH ORDER
%

\subsection{Zeroth order peak-time}\label{ZerothOrderPeakTime}
Without pileup, a single pulse at time $0$ has a voltage signal that can be expressed as
\begin{equation}
V(t)\ = \ v_0 f(t)  
\end{equation}
where $f(t)$ is the normalized pulse shape, such that at its maximum, $f = 1$.  Therefore the peak voltage is $V = v_0$.  The peak-time is given by the first solution of
\begin{equation}
\frac{df(t)}{dt} \Bigr \rvert_{t^0_p}  = 0
\end{equation}
and we will denote it by the symbol $t^0_p$.  Thus, 
\begin{dgroup*}
\begin{dmath*} f^{\prime}(t^0_p)=0 \end{dmath*}
\begin{dmath*} f(t^0_p) = 1 \end{dmath*}
\begin{dmath*} V(t^0_p) = v_0 \end{dmath*}
\end{dgroup*}
%
% FIRST ORDER
%
\subsection{First order peak pileup}

\subsubsection{First order peak-time}

First order peak-pileup is depicted in figure \ref{fig:ThreePanel1}(a) and figure \ref{fig:SummedPeak1}.  The apparent pulse is a linear combination of unit pulses $f(t)$, 
\begin{equation}\label{eq:1stOrderSignal}
V(t)\ = \ v_0 f(t - t_0) + v_1 f(t-t_1) 
\end{equation}
where $t_0, t_1$ are the incident event times, and $v_0, v_1$ are their peak amplitudes.  For the constant rate process, the total time offset of the pair is irrelevant, so we can let $t_0 = 0$.  Then $t_1$ is equal to the separation of the two events, $s_1$.  The maxima of equation \eqref{eq:1stOrderSignal} are given by 
\begin{equation}\label{eq:1stOrderPeakGeneral}
\frac{d}{dt} V(t)  \Bigr \rvert_{t^1_p} =\ \Biggl [ v_0 \frac{d f(t)}{dt}  + v_1  \frac{d f(t - s_1)}{dt}\Biggr ]_{t^1_p} = 0
\end{equation}
We define the symbol $t^1_p$ as the position of the measured first-order peak, such that it satisfies the above equation.  As discussed in section \ref{sec:PPU} the observed peak is shifted forward, so $t^1_p \geq t^0_p$, and it depends on the random variables  $s_1,v_0, v_1$:
\begin{equation}
 t^1_p =  t^1_p(s_1,v_0, v_1)
 \end{equation}
Having a model for $t^1_p(s_1,v_0, v_1)$ is necessary so that the sum \eqref{eq:1stOrderSignal} can be evaluated for the observed first-order energy.  Depending on the pulse shape $f(t)$, equation \eqref{eq:1stOrderPeakGeneral} might have a closed form solution for  $t^1_p$.   In previous work simplifications are employed such that $t^1_p$ is given by simple geometrical arguments.  But in the present case, substitution of the pulse shape, equation \eqref{eq:PulseShape}, into equation \eqref{eq:1stOrderPeakGeneral} results in an expression that cannot be readily inverted.   As a result we use an empirical model for the peak-time that is presented in section \ref{GBMPeakTime}.

Note that for a given set of values $(s_1,v_0, v_1)$ there may be two solutions to equation \eqref{eq:1stOrderPeakGeneral}  (two maxima in the peak interval, for example in figure \ref{fig:SummedPeak1}).  Selecting the correct peak requires knowledge of the peak-finding algorithm of the detector system in question.  The buffering scheme of the GBM pulse-height analyzers generally causes the last maximum to be the measured one.  However if $v_0$ is much larger than $v_1$, and $s_1$ is in the buffer, $t^1_p \rightarrow t^0_p$ because the falling derivative has a much larger (negative) value than the rising derivative in this region.  When the two add, the net change is still negative, so the decreasing sample criterion is satisfied.  Such discrete logic is expressed as piecewise behavior in the model for $t^1_p$.

%%%%%%%%%%%%%%%%%%%%%
%       First ORDER PEAK SECTION
%%%%%%%%%%%%%%%%%%%%%
\subsubsection{First order peak-energy}
Let us assume that we have a sufficient model for $t^1_p(s_1,v_0, v_1)$.  Unlike $f(t^0_p)=1$, $f(t^1_p) \neq 1$ in general.  The recorded pulse height is
\begin{equation}
V(t^1_p) = v_0 f(t^1_p) + v_1 f(t^1_p-s_1)
\end{equation}
and the recorded first-order pileup energy  $\varepsilon_1$ is 
\begin{equation}
\varepsilon_1 = \xi[ v_0 f(t^1_p) + v_1 f(t^1_p-s_1) ]
\end{equation}
where $\xi[v]$ is the \textit{channel} or \textit{voltage-to-energy} conversion, determined by standard methods of instrument calibration.  For modeling input spectra, we assume the inverse $\xi^{-1}$ exists and the coefficients of input pulses can be calculated as $v_i =  \xi^{-1}[E_i]$ where $E_i$ is energy deposited by a detected gamma-ray.  $E_i$ is of course converted from incident photon energy by the various physical processes of the detector, but this is extraneous to the current problem.  It is sufficient to say that $E_i$ is the recorded energy of the $i^{th}$ count in the absence of pileup effects.
 
\begin{figure*} [t]
\centering
\includegraphics[width=0.7\linewidth]{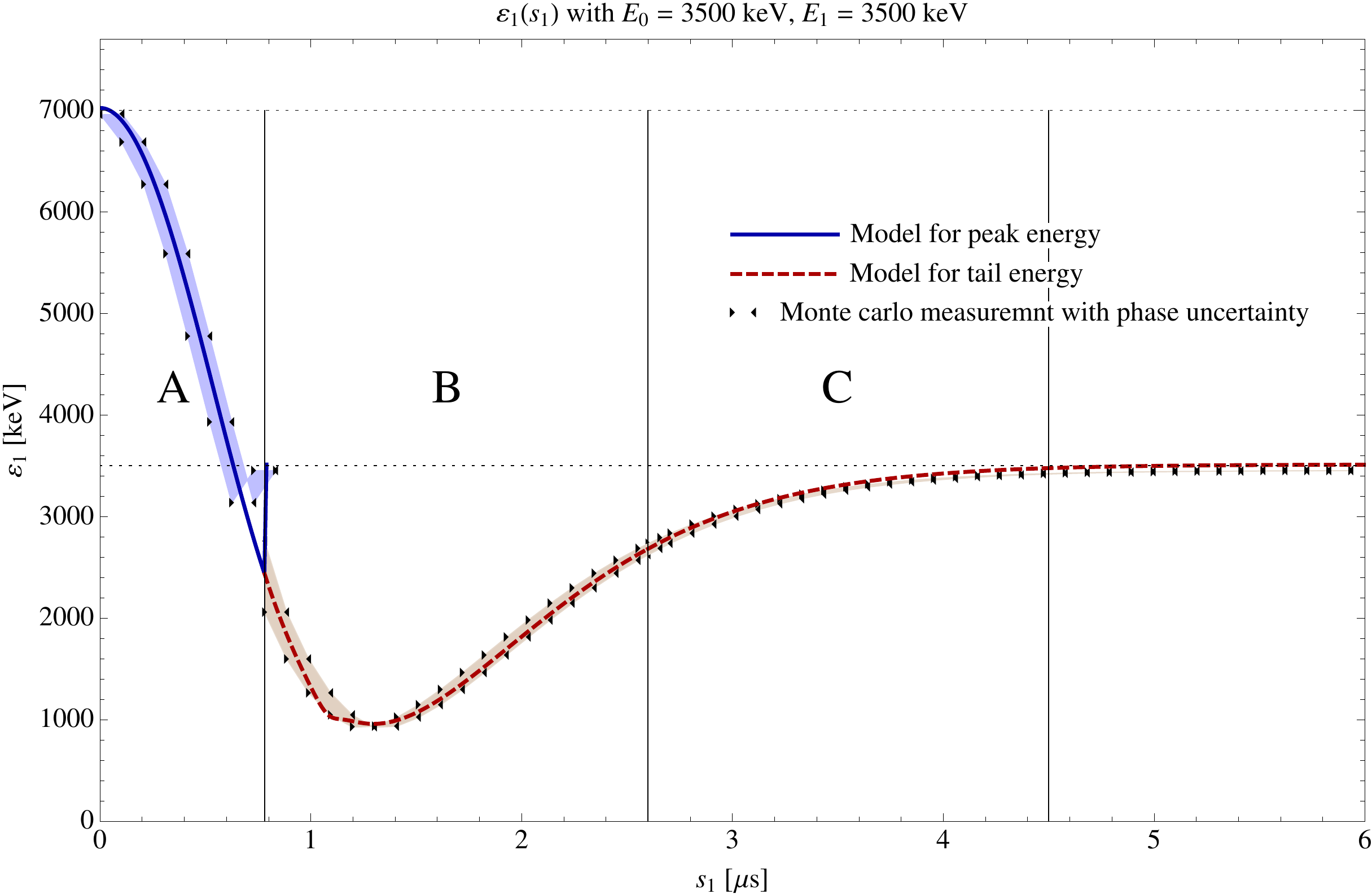}
\caption{Plot of 1st-order recorded energy $\varepsilon_1( s_1,E_0, E_1)$ for $E_0, E_1$ at 3500 keV, if the second event occurs at $s_1$.  The curves represent the model for measured energy using the function $t_p^1$, which is different in the peak than in the tail.  Shaded areas between points give the range of recorded energy due to the phasing of digital samples. Discontinuity in the peak curve (BLUE) occurs when the separation is large enough for the initial pulse height to be measured correctly.  The divisions A, B, C refer to the pulse partition of section \ref{PulsePartitionSection}. The curve in A refers to a single measurement of two events that have pulse pileup.  The curves in B-C refer to measurement of a second event (tail effect), which is addressed in section \ref{TailEffect}.  In the latter case, \emph{two} measurements are made, with peak zero measured at the input value of $E_0$.  }
\label{fig:MeasuredEnergy1}
\end{figure*}

The recorded 1st-order energy, $\varepsilon_1$, for a given instance of the variables $(s_1,E_0, E_1)$, is:
 %%%%
% EQUATION:    First order peak energy
%	 Er  or   E_1
%%%%%
\begin{dmath} \label{eq:1stOrderEnergy}
\varepsilon_1= \xi \biggl[  \xi^{-1}[E_0] f(t^1_p) + \xi^{-1}[E_1] f(t^1_p-s_1)  \biggr]  \equiv \varepsilon_1( s_1,E_0, E_1)
\end{dmath}
% \begin{equation}\label{eq:1stOrderEnergy}
%\boxed{ \varepsilon_1( s_1,E_0, E_1) = \xi \biggl[  \xi^{-1}[E_0] f(t^1_p) + \xi^{-1}[E_1] f(t^1_p-s_1) \biggr] }
% \end{equation}
%
Note that if $\xi[v]$ is approximately of the form $ y = mx$, then this simplifies to $ E_0 f(t^1_p) + E_1 f(t^1_p-s_1)$.  In section \ref{GBMPeakTime} we give the expressions for $t_p^1$ and $\varepsilon_1$ using the true GBM pulse shape.  Figure \ref{fig:MeasuredEnergy1} is calculated with the above formula, and $t_p^1$ from that section.  Discontinuity at the end of region A occurs because the two pulses are sufficiently separated for the first to be measured correctly.  It is expressed as piecewise behavior in $t_p^1$, given in section \ref{GBMPeakTime}.

%%%%
% FIGURE:  6
%	Er
%%%%%

%%%%
% FIGURE:  7
%	PPU Likelihood
%%%%%
\begin{figure*}[t]
\centering
\includegraphics[width=\linewidth]{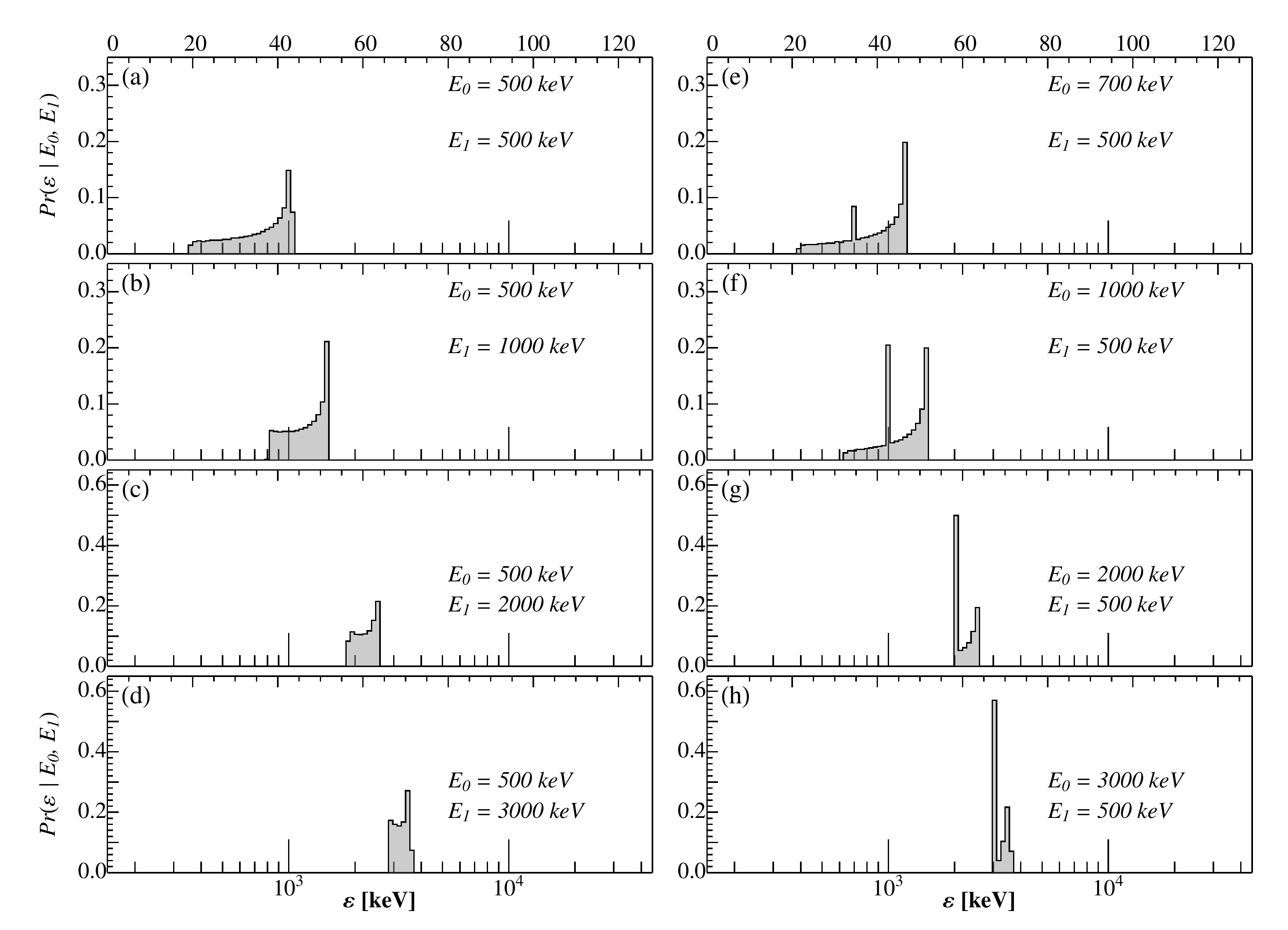}
\caption{$\text{Pr}^1_{\rm peak}(\varepsilon | E_0, E_1 )$, for several combinations of $E_0, E_1$.  Assuming two peak pileup events, these recorded energy distributions result from varying their separation.  When $E_1 \geq E_0$ (panels (a) - (d)), peak probability occurs at $\varepsilon = E_0 + E_1$, indicating this is the most frequent case.  The width of each distribution demonstrates that $E_0 + E_1$ is not a good approximation for $\varepsilon$ in general.  A second peak, appearing in panels (e) - (h), indicates that $E_0$ has a higher probability of being distinguished as it becomes larger than $E_1$.  }
\label{fig:PeakPPULikelihood}
\end{figure*}

%%%%%%%%
%
%	Recorded energy distribution
\subsubsection{Probability distribution of $\varepsilon_1$}\label{1stOrderPeakSpecDerive}
What we need is a model giving the observed distribution of the recorded energy $\varepsilon_1$ in the event of peak pileup.   This means we need the distribution of $\varepsilon_1$ over all possible realizations of $\{(s_1,E_0, E_1)\}$.  Let us examine the distribution $\varepsilon_1$ due to $s_1$ \emph{only}, holding $E_0,E_1$ fixed at some arbitrary values.  This gives a conditional form of equation \eqref{eq:1stOrderEnergy},  $\varepsilon_1( s_1\ |\ E_0, E_1)$. 

%For a general function $Y(X)$ of a random variable $X$, $Y$ is also a random variable, and has the distribution (PDF) $f_Y(y)$.  If $Y(X)$ is a monotonic function over some interval of $X$, and $f_X(x)$ is the PDF of $X$ in this interval, then $f_Y(y)$ can be written
%\begin{equation}
%f_Y(y) = f_X(x) \lvert \frac{dX}{dY} \rvert = \frac{ f_X(x) } {  \lvert \frac{dY}{dX} \rvert }
%\end{equation}
%and the argument $x$ can be replaced by means of the inverse $x = Y^{-1}(y)$. \cite{Grimmet}

Given the interval PDF $f_{s|n=1}(s_1)$ of equation \eqref{eq:nthOrderPeakIntervalDist}, the PDF of the dependent variable $\varepsilon_1$ can be found in the standard way \cite{Grimmet}.  For $s_1$ in the peak region, $\varepsilon_1( s_1\ |\ E_0, E_1)$ is monotonic, and we can write its probability density as

\begin{equation}
f_{\rm peak}^{\langle 1 \rangle }(\ \varepsilon_1(s_1)\ ) =  \frac{f_{s|n=1}(s_1)}{\bigl\lvert \frac{\partial \varepsilon_1(s)}{\partial s}  \bigr\rvert}_{s=s_1}
 \end{equation}
 where $\varepsilon_1( s_1\ |\ E_0, E_1)$ is written $\varepsilon_1(s_1)$ for brevity.  It is understood that $t^1_p$ and $\varepsilon_1$ are functions of the three random variables.
 
 We can write the probability of pulse pileup into an energy interval $\varepsilon_1(s_1)$ to $\varepsilon_1(s_1) + d\varepsilon_1$ as
\begin{align}
d[\text{Pr}_{\rm peak}^{\langle 1 \rangle }{(\varepsilon_1(s_1) \ |\ E_{0}, E_{1})}] =& f_{\rm peak}^{\langle 1 \rangle }(\varepsilon_1) d\varepsilon_1
 \end{align}
 Then we can write the following for the probability that $\varepsilon_1$ is in the discrete channel $\varepsilon$ to $\varepsilon + \Delta\varepsilon$, in the event that it is due to $E_0, E_1$ pileup:

\begin{align}
\text{Pr}_{\rm peak}^{ \langle 1 \rangle }{(\varepsilon_1 \ |\ E_{0}, E_{1})} =&  \int \limits_{\varepsilon}^{\varepsilon + \Delta\varepsilon}  f_{\rm peak}^{\langle 1 \rangle }(\varepsilon^{\prime}) d\varepsilon^{\prime} \\
=&  \int \limits_{\varepsilon}^{\varepsilon + \Delta\varepsilon}   \frac{ f_{s|n=1}(s_1^{\prime})}{\bigl\lvert \frac{\partial \varepsilon_1(s)}{\partial s}  \bigr\rvert}_{s=s_1^{\prime}} d\varepsilon^{\prime}
\end{align}
This can be converted to an integral over $s_1$ using the positive-definite Jacobian determinant (probability must be positive)
\begin{equation}
\biggl\lvert \frac{ \partial( \varepsilon^{\prime} )} { \partial(s_1) }  \biggr\rvert = {\biggl\lvert \frac{\partial \varepsilon_1(s)}{\partial s}  \biggr\rvert}_{s=s_1}
\end{equation} 
giving
%
%%%%
%  EQUATION
%	1st order kernel
%
%\begin{framed}
\begin{dgroup*}
\begin{dmath*}
\text{Pr}_{\rm peak}^{ \langle 1 \rangle }{(\varepsilon_1 \ |\ E_{0}, E_{1})} 
\end{dmath*}
\begin{dmath*} 
=  \int \limits_{s(\varepsilon)}^{s(\varepsilon + \Delta\varepsilon)}    \frac{f_{s|1}(s_1^{\prime})}{\bigl\lvert \frac{\partial \varepsilon_1(s)}{\partial s} \bigr\rvert}_{s_1^{\prime}}   {\biggl\lvert \frac{\partial \varepsilon_1(s)}{\partial s}  \biggr\rvert}_{s_1^{\prime}}       ds_1^{\prime} 
\end{dmath*}
\begin{dmath}   \label{eq:1stOrderProbTensor}
= \int \limits_{s(\varepsilon)}^{s(\varepsilon + \Delta\varepsilon)}  f_{s|1}(s_1^{\prime})    ds_1^{\prime}
\end{dmath}
\end{dgroup*}
%\end{framed}
%
%
%%
% PPU  pdf plot
%%
%\begin{figure}[ht]
%\centering
%\epsfig{file=FpeakPDF.eps, keepaspectratio=true, width=\linewidth}
%\caption{$f_{\rm peak}^{ \langle 1 \rangle }{(\varepsilon_1 \ |\ E_{0}, E_{1})}$ for $E_0, E_1$ at 3500 keV.}
%\label{fig:fpeakPDF}
%\end{figure}

%%%%%%%%%
% FIGURE:
% PPU map 
%%%%
%\begin{figure*}[t!]
%\centering
%\epsfig{file=PPUProbMap.eps, keepaspectratio=true,  width=4.0in}
%\caption{$\text{Pr}_{\rm peak}^{ \langle 1 \rangle }{(\varepsilon_1 \ |\ E_{0}, E_{1})}$ for $\varepsilon_1$ in channel 70 (vertical line).  This is the probability of recording 1st-order peak pileup in this channel, as a function of $(E_0, E_1)$.  For each pixel $I^{ \langle 1 \rangle }(s_1)$ has been integrated over a range $\Delta s$ corresponding to the measurement bin width, $\Delta E$, giving the total probability in the pixel.  For a given $(s_1,E_{0}, E_{1})$ then, $\Delta s$ satisfies $\lvert \varepsilon_1(s_1) - \varepsilon_1(s_1 + \Delta s) \rvert \leq \Delta E$.  White pixels correspond to Pr$=0$.  A black pixel is Pr$=1$, meaning that $(E_0, E_1)$ are always recorded in this channel, regardless of $s_1$.  The feature at $E_1 \sim 3.5$ MeV is due to the discrete logic of the instrument's peak finding algorithm, which we have modeled in the expression for $t^1_p$. Note that this is a demonstration map and the energy channel edges are not the exact GBM edges.}
%\label{fig:PPUProbMap}
%\end{figure*}

Examples of this likelihood function are shown in figure \ref{fig:PeakPPULikelihood}.  Evaluation of the above integration limits evidently requires knowledge of an inverse function giving the separation in terms of the summed peak and two energies.  Therefore, if a closed form expression for equation \eqref{eq:1stOrderProbTensor} is desired, the recorded energy $\varepsilon_1(s_1 \ |\ E_0, E_1)$ must be an analytical expression and have an inverse, $s_1 = \varepsilon_1^{-1}(\varepsilon, E_0, E_1)$.  

The simplification of \cite{taguchi:2010} using triangular pulses allows such an inversion.  In the present case, the true pulse shape results in an expression for $\varepsilon_1$ which is too complicated to invert.  Thus we have the pileup likelihood expressed in implicit form, and must evaluate it numerically.  In section \ref{Evaluation} we describe a simple algorithm for doing the implicit evaluation.
%

%%%%%
% EQUATION	
%	1st Order correction
Regardless, we can write the first-order peak pileup contribution to the spectrum as
\begin{dmath} \label{eq:1stOrderCorrection}
P_{ \langle 1 \rangle }( \varepsilon_1 ) = \int \limits_{0}^{\infty} \int \limits_{0}^{\infty}  \text{Pr}_{\rm peak}^{ \langle 1 \rangle }{(\varepsilon_1 \ |\ E_{0}, E_{1})} \times S(E_{0})S(E_{1}) dE_{0} dE_{1}
\end{dmath}
$S(E)$ is the PDF of the detected energy spectrum, and is normalized to unity.  $S(E)$ is typically modeled in terms of an externally incident photon spectrum, $S_{\gamma}(E_{\gamma})$, and the detector response function $R(E | E_{\gamma})$, for example as described in \cite{Knoll}:
\begin{equation}
S(E) = \int R(E | E_{\gamma}) S_{\gamma}(E_{\gamma}) dE_{\gamma}
\label{eq:ResponseConvolution}
\end{equation}
where $S_{\gamma}$ is the PDF of the photon spectrum.  Since real detector systems are sensitive over a finite energy range, we note that the limits of integration in equation \eqref{eq:1stOrderCorrection} can be finite values $E_{\rm min} \rightarrow E_{\rm max}$ as long as $S(E)$ is normalized over this interval.  The PDF of the spectral contribution is
\begin{equation}
p_{ \langle 1 \rangle }( \varepsilon_1 ) = \frac{\partial}{\partial  \varepsilon_1 } P_{ \langle 1 \rangle }( \varepsilon_1 )\label{eq:1stOrderCorrectionPDF}
\end{equation}

%%%%%%%%%%%%%%%%%%%%%
%       HIGHER ORDER PEAK SECTION
%%%%%%%%%%%%%%%%%%%%%
\subsection{Higher order peak terms}\label{nthOrderPeakSpecDerive}

For $n^{th}$ order pileup we require an expression giving the recorded energy within the peak interval.  The $n+1$ pulses will be superimposed and recorded as a single pulse height $\varepsilon_n$.  This energy would generally be dependent on the energies and separations of the piled-up events:
\begin{align*}
\varepsilon_n\ &=\ \xi[\sum_{i=0}^n v_i f(t-t_i)]\\
\Rightarrow\ \varepsilon_n &= \varepsilon_n( s_1, \dotsc , s_n\ |\ E_0, E_1, E_2, \dotsc, E_n )\\
\text{for}\ s_i &=\ t_{i} - t_{i-1}
\end{align*}
The other expressions of the previous section would generalize in a similar way, leading to the $n^{th}$ order peak correction 
\begin{dmath}
P_{ \langle n \rangle }( \varepsilon_n ) = \int dE_0  \int dE_1  \dotsi  \int dE_n\times\  \text{Pr}_{\rm peak}^{ \langle n \rangle }(\varepsilon_n\ |\ E_0, \dotsc, E_n )S(E_0)\dotsm S(E_n)  
\end{dmath}
These expressions are complicated and unwieldy, so instead we use the iterative approximation of \cite{taguchi:2010}.  This scheme approximates higher order variations due to the added random variables by using results from the previous order. 

To calculate the $n^{th}$ order correction, the previous order term  $p_{ \langle n-1 \rangle }$ is used:
%%
% Eqn:  Nth Order Correction
%%
\begin{framed}
\begin{dmath} 
P_{ \langle n \rangle }( \varepsilon_n ) = \int \limits_{0}^{\infty} \int \limits_{0}^{\infty}   \text{Pr}_{\rm peak}^{ \langle n \rangle }(\varepsilon_n \ |\ E_{n-1}, E_{n}) \times\  p_{ \langle n-1 \rangle }(E_{n-1})S(E_{n}) dE_{n-1} dE_{n} 
 \label{eq:NthOrderCorrection}
\end{dmath}
\end{framed}
with the PDF 
\begin{equation}
\boxed{    p_{ \langle n \rangle }( \varepsilon_n ) = \frac{\partial}{\partial  \varepsilon_n } P_{ \langle n \rangle }( \varepsilon_n )    }
\end{equation}
and the kernel $\text{Pr}_{\rm peak}^{ \langle n \rangle }$ is evaluated using the $n^{th}$ order interval statistics:
%nth Order kernel
\begin{equation}\label{eq:NthOrderProbTensor}
\boxed{ \text{Pr}_{\rm peak}^{ \langle n \rangle }{(\varepsilon_n \ |\ E_{n-1}, E_{n})} =  \int \limits_{s(\varepsilon)}^{s(\varepsilon + \Delta\varepsilon)}  f_{s|n}(s^{\prime})    ds^{\prime} }
\end{equation}
We make the approximation that $\varepsilon_n \approx \varepsilon_n(s_n \ |\ E_{n-1}, E_{n})$ and has the same form as $\varepsilon_1(s_n \ |\ E_{n-1}, E_{n})$.  $f_{s|n}(s)$ is the interval distribution for $n$ events in the peak, equation \eqref{eq:nthOrderPeakIntervalDist}.

We can unify the $1^{st}$ order correction with \eqref{eq:NthOrderCorrection} by defining $p_{ \langle 0 \rangle }(E) = S(E)$.  Then equation  \eqref{eq:NthOrderCorrection} reduces to \eqref{eq:1stOrderCorrection} for $n=1$.
%

%%%%%%%%%
%
%  TAIL EFFECT
%
%%%%%%%%%
\section{Tail effect}\label{TailEffect}
The ``tail effect" is spectral distortion caused by the bipolar pulse tail, which reduces pulse heights occurring before baseline recovery.  In addition to energy-lowering distortion, energy-dependent losses can occur if the reduced peak is below a lower-level threshold.  The model of \cite{taguchi:2010} presents a clear and relatively accurate model for tail effects in the output spectrum.  However, in their method there is not the intervening deadtime region (interval `B').  For modeling purposes this means tail subtraction effects are taken to depend only on the peak state.  In the present context we must consider the events in the peak \emph{and} deadtime, since pulses in either region contribute to a negative tail when the instrument again becomes live.  The method of \cite{taguchi:2010} uses further simplifications by reducing the number of random variables: input tail events are assumed to be uniformly distributed in time and have the same energy, which is taken to be the mean energy from the modeled input spectrum; i.e., $S(E) \rightarrow \delta(E- \langle E \rangle)$.

We present an alternative tail technique which deals with the intervening deadtime interval (region `B' in figures \ref{fig:UnitPulseIntervals} and \ref{fig:ThreePanel2}), and models energy and timing variations of tail pileup events.  This method is similar to that used for peak pileup in that pileup likelihood functions of the form of equation \eqref{eq:1stOrderProbTensor} are derived for measurement in the tail region (region `C').  Measurement in this region is sensitive to the previous two, so the likelihood scheme is more complex than in the peak case.  The number of events in A and B affect the C measurement because their negative tails combine (figure \ref{fig:ThreePanel2}(c)).  Additionally, peak pileup in C can occur (figure \ref{fig:ThreePanel2}(b)).  The resulting pulse heights in the latter case are modified by both peak \emph{and} tail effects.  We account for the various possibilities by isolating an `A$+$C' effect, which is the case of pulses in A or C only (or both), and a `B$+$C' effect, which assumes no zeroth event and pulses only in B or C (or both).  For each case a likelihood function for the tail pileup energy is calculated.  The spectrum of the total pulse configuration, with `A$+$B$+$C' dependence, is calculated by convolving the corresponding `A+C' component with the `B+C' likelihood.

%An advantage of this technique is that it represents an approximate treatment of paralyzed deadtime effects, in which a train of nearby pulses (each one within $\tau_A$ of the previous) delays registration of the summed %peak and thus extends the deadtime.  

%
% Er Tail
%
\subsection{Tail energy} \label{TailEnergy}
In region A we employed the function $t^1_p(s_1,v_0,v_1)$ to give the time of the first-order peak.  For tail pileup we make the simplifying assumption that the location of the summed peak, were it to be measured in C, is approximately at the peak of the second input pulse, i.e., $t^1_p \approx t^0_p + s_1$.  Thus equation \eqref{eq:1stOrderEnergy} becomes
%%%%
%  EQUATION :   First order TAIL energy
%	Er Tail
\begin{dmath} \label{eq:1stOrderTailEnergy}
\varepsilon_1\approx \xi \biggl[  \xi^{-1}[E_0] f(t^0_p + s_1) + \xi^{-1}[E_1] f(t^0_p) \biggr]   \equiv \varepsilon_1( s_1,E_0, E_1) \condition{for tail energies}
\end{dmath}

%\begin{equation}\label{eq:1stOrderTailEnergy}
%\boxed{ \varepsilon_1( s_1\ |\ E_0, E_1) \approx \xi \biggl[  \xi^{-1}[E_0] f(t^0_p + s_1) + \xi^{-1}[E_1] f(t^0_p) \biggr]   \text{   (Tail energy)}  }
%\end{equation}
For modeling the second tail effect, we also need an expression for peaks measured in B.  This is given in equation \eqref{eq:TotalPeakTime}, which is a smooth joining of the complicated function for region A peaks, and the simplified expression for region C.  Figure \ref{fig:MeasuredEnergy1} shows that this method is reasonably accurate.

%
%  FIGURE:  8
%	Tail Effect LIKELIHOOD
\begin{figure*}[t]
\centering
\includegraphics[width=\linewidth]{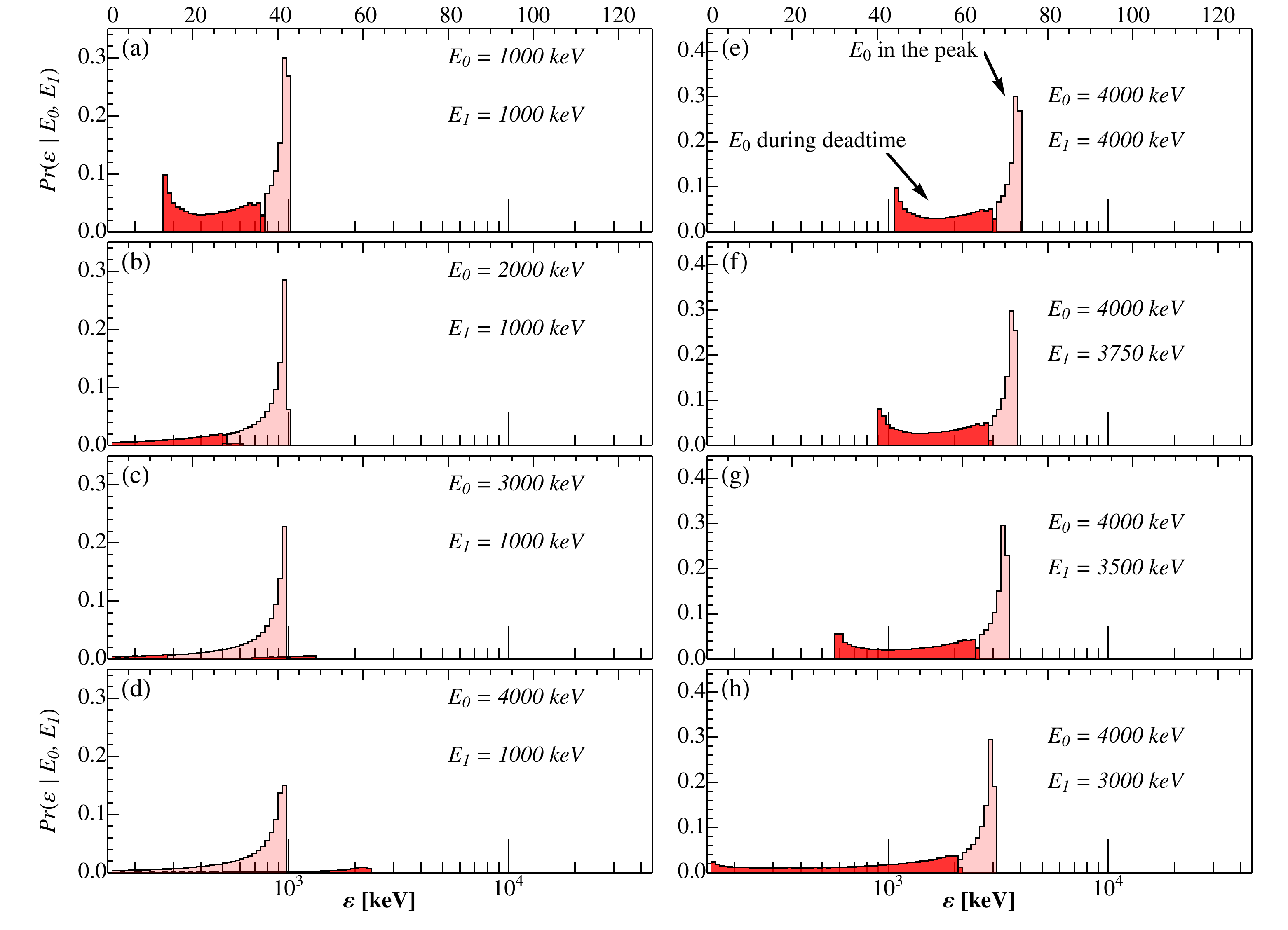}
\caption{Recorded energy distributions for the two tail effects.  In each plot the lighter histogram plots $\text{Pr}^1_{A+C}(\varepsilon | E_0, E_1 )$, which assumes $E_0$ is the peak energy, and the separation $s_1$ ranges from $\tau_A + \tau_B$ to $\tau_A + \tau_B + \tau_C$.  The darker histogram plots $\text{Pr}^1_{B+C}(\varepsilon | E_0, E_1 )$, which assumes $E_0$ is in the dead region (B), and $s_1$ varies from $\tau_B$ to $\tau_B+\tau_C$.   }
\label{fig:TailPPULikelihood}
\end{figure*}

%%%%%
%
% TAIL SPECTRUM 
%
%%%%%

\subsection{Recorded spectrum of tail events}\label{TotalTailSpectrum}
The recorded tail spectrum depends on the total pulse the state $\langle a,b,c \rangle$.  We define the following symbols to represent the spectrum of measured tail energies associated with a window state:

\begin{table}[!h]
\centering
\begin{tabular}{c p{0.5\linewidth}}
\hline\\
$Q_{\langle a,b,c \rangle}(\varepsilon) \equiv$  & probability of recording energy $\varepsilon$ in the tail \\
$q_{\langle a,b,c \rangle}(\varepsilon) \equiv$ & PDF of recorded energy, $\frac{\partial}{\partial\varepsilon} Q_{\langle a,b,c \rangle}$\\
\\
\hline
\end{tabular}
\caption{Symbols used for the distribution and PDF of recorded energy of the tail events in a state}
\end{table}
The effect of events \emph{only} in A upon measurements in C is first expressed in the $\langle a,0,c \rangle$ components.  For $c=0$ there is no tail count and thus no tail measurement, so such terms are zero.  For $a + c > 1$ an iterative scheme is used.  In general, a succession of terms will be used to go from first order tail pileup to higher orders: $Q_{\langle 0,0,1 \rangle} \rightarrow Q_{\langle 0,0,c \rangle} \rightarrow Q_{\langle a,0,c \rangle} \rightarrow Q_{\langle a,b,c \rangle}$.

%%%%%%%%%%%
%
% Subsections:
%	First tail effect
%
\subsubsection{First tail effect, $Q_{\langle a,0,c \rangle}(\varepsilon)$}

% FIGURE 9:  Tail peak
%
\begin{figure} [ht!]
\centering
\includegraphics[width=\linewidth]{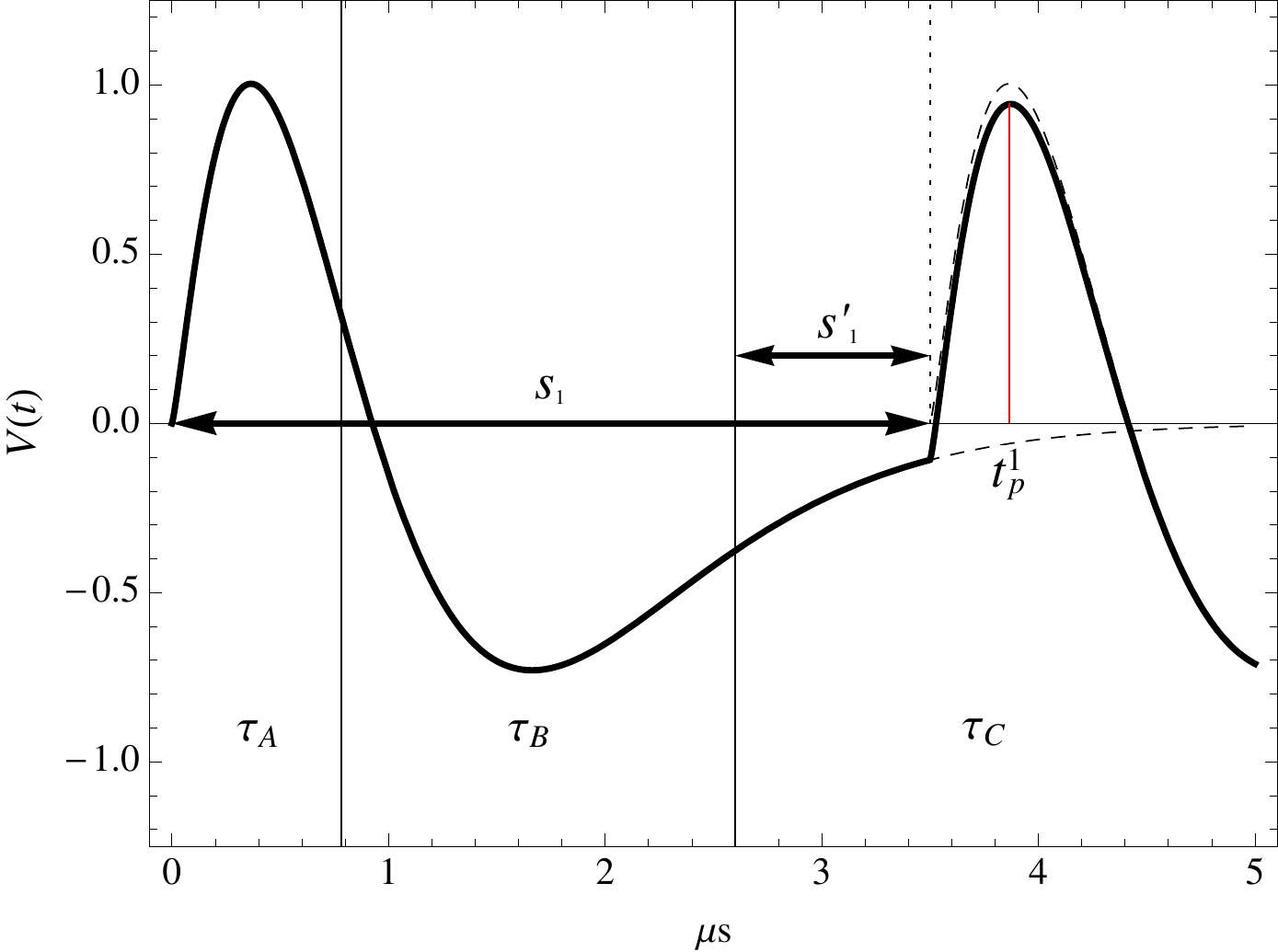}
\caption{$\langle 0,0,1 \rangle$ pileup event.  $t^1_p \approx t^0_p + s_1$ for the tail pulse.  $s_1^{\prime}$ is distributed according to equation \eqref{TailIntervalDist}.}
\label{fig:ACPulseIntervals}
\end{figure}

We begin by calculating the likelihood of recording energy $\varepsilon$ in region C, in the event of tail pileup.  Here we assume a zeroth event to define the pulse window, and additional events in A and C.  For first order, $a=0$ and $c=1$.  Figure \ref{fig:ACPulseIntervals} depicts this configuration.

Recall that the interval PDF for a single tail event is $f_{s|c=1}(s_1^{\prime}) = \frac{1}{\tau_C}$,  where  $s_1^{\prime} = s_1 - (\tau_A + \tau_B)$ (equation \eqref{TailIntervalDist}).  However the full separation $s_1$ is used to calculate the measured energy using equation \eqref{eq:1stOrderTailEnergy}.  Assuming $E_0$ in $\tau_A$, $E_1$ in $\tau_C$, the probability of tail pileup into the discrete energy bin $[\varepsilon, \varepsilon + \Delta \varepsilon]$ is then:
%
% Equation:  Pr_tail^1
\begin{equation}
\text{Pr}_{A+C}^{ \langle 1 \rangle }{(\varepsilon \ |\ E_{0}, E_{1})} =  \int \limits_{s(\varepsilon)}^{s(\varepsilon + \Delta\varepsilon)}  f_{s|c=1}(s_1^{\prime})    ds_1^{\prime} 
\end{equation}
and the $1^{st}$ order spectral component is
%
% Equation: Q_001 %
\begin{dmath} \label{eq:1stOrderTailCorr}
Q_{\langle 0,0,1 \rangle} (\varepsilon) = \int \limits_{0}^{\infty} \int \limits_{0}^{\infty}  \text{Pr}_{A+C}^{ \langle 1 \rangle }(\varepsilon \ |\ E_{0}, E_{1}) \times  S(E_{0})S(E_{1}) dE_{0} dE_{1} 
\end{dmath}
For $\langle a,0,1 \rangle$ states we approximate the effect of the multiple orders in A by using $p_{\langle a \rangle}$ in the integral, and using the $\text{Pr}_{A+C}^{ \langle 1 \rangle }$ kernel, since the interval distribution in C does not depend on the number of events in A (section \ref{MultipleIntervals}).  The spectrum measured in the tail for these configurations is
%
% Equation: Q_a01 %
\begin{framed}
For $a \geq 0,\ b=0, \ c = 1$,
\begin{dmath}\label{eq:TailCorra01}
Q_{\langle a,0,1 \rangle}(\varepsilon) = \int \limits_{0}^{\infty} \int \limits_{0}^{\infty}  \text{Pr}_{A+C}^{ \langle 1 \rangle }(\varepsilon \ |\ E^{\prime}, E^{\prime\prime}) \times  p_{\langle a \rangle}(E^{\prime})S(E^{\prime\prime}) dE^{\prime} dE^{\prime\prime}
\end{dmath}
\end{framed}
For $a= 0$ this reduces to equation \eqref{eq:1stOrderTailCorr} since $p_{\langle 0 \rangle}(E) = S(E)$.  For zero events in the tail, there can be no recorded tail energy, so $q_{\langle a,b,0 \rangle}(E) = 0$ for all $a,b$.

For $\langle 0,0,c \rangle$ states and $c > 1$, we do an iterative approximation using the previous order.  However, there are now multiple pulses in the region C, and we approximate their measured energy as a case of $(c-1)^{th}$ order peak pileup.  But now the primary and additional event have a negative baseline, and so we approximate them using $q_{\langle 0,0,c-1 \rangle}$ as the primary peak distribution\footnote{Much like the $a^{th}$ order peak iteration uses $p_{a-1}(\varepsilon)$ in the primary position.}, and $q_{\langle 0,0,1 \rangle}$ as the spectrum of additional events.  Then we use the kernel $\text{Pr}_{\rm peak}^{\langle c-1 \rangle}$ from equation \eqref{eq:NthOrderProbTensor}.  This is an approximation since $\text{Pr}_{\rm peak}^{\langle n \rangle}$ models pileup in the interval $\tau_A$ instead of $\tau_C$.  However multiple events in $\tau_C$ tend to be clustered due to the distribution $I^{\langle n \rangle}$, and so the small difference between interval sizes $\tau_A$ and $\tau_C$ is not a major source of error.  Then the $\langle 0,0,c \rangle$ tail spectrum is
%
% Equation: Q_00c %
\begin{framed}
For $a=0,\ b=0,\ c > 1$,
\begin{dmath} \label{eq:TailCorr00c}
Q_{\langle 0,0,c \rangle}(\varepsilon) = \int \limits_{0}^{\infty} \int \limits_{0}^{\infty}  \text{Pr}_{\rm peak}^{ \langle c-1 \rangle }(\varepsilon \ |\ E^{\prime}, E^{\prime\prime}) \times  q_{\langle 0,0,c-1 \rangle}(E^{\prime})q_{\langle 0,0,1 \rangle}(E^{\prime\prime}) dE^{\prime} dE^{\prime\prime}
\end{dmath}
\end{framed}

Finally, for $\langle a,0,c \rangle$ states, we use the $a^{th}$ order peak spectrum with the $q_{\langle 0,0,c \rangle}(\varepsilon)$ just calculated:
%
%Equation: Q_a0c
\begin{framed}
For $a > 0,\ b=0,\  c > 1$,
\begin{dmath}\label{eq:TailCorra0c}
Q_{\langle a,0,c \rangle}(\varepsilon) = \int \limits_{0}^{\infty} \int \limits_{0}^{\infty}  \text{Pr}_{A+C}^{ \langle c \rangle }(\varepsilon \ |\ E^{\prime}, E^{\prime\prime}) \times  p_{\langle a \rangle}(E^{\prime})q_{\langle 0,0,c-1 \rangle}(E^{\prime\prime}) dE^{\prime} dE^{\prime\prime}
\end{dmath}
\end{framed}
where the $c^{th}$ order kernel is approximated as
\begin{equation}
\text{Pr}_{A+C}^{ \langle c \rangle }{(\varepsilon \ |\ E_{a}, E_{c})} =  \int \limits_{s(\varepsilon)}^{s(\varepsilon + \Delta\varepsilon)} f_{s|c}(s_1^{\prime})    ds_1^{\prime} 
\end{equation}
and the recorded energy is equation \eqref{eq:1stOrderTailEnergy} with $E_a, E_c$ replacing $E_0,E_1$ and  $f_{s|c}$ is the tail interval density, equation \eqref{TailIntervalDist}.

%%%%%%%%%%%%%%
% 
% Second Tail component
%
%%%%%%%%%%%%%%

% FIGURE 10:  B-C peak
%
\begin{figure} [ht!]
\centering
\includegraphics[width=\linewidth]{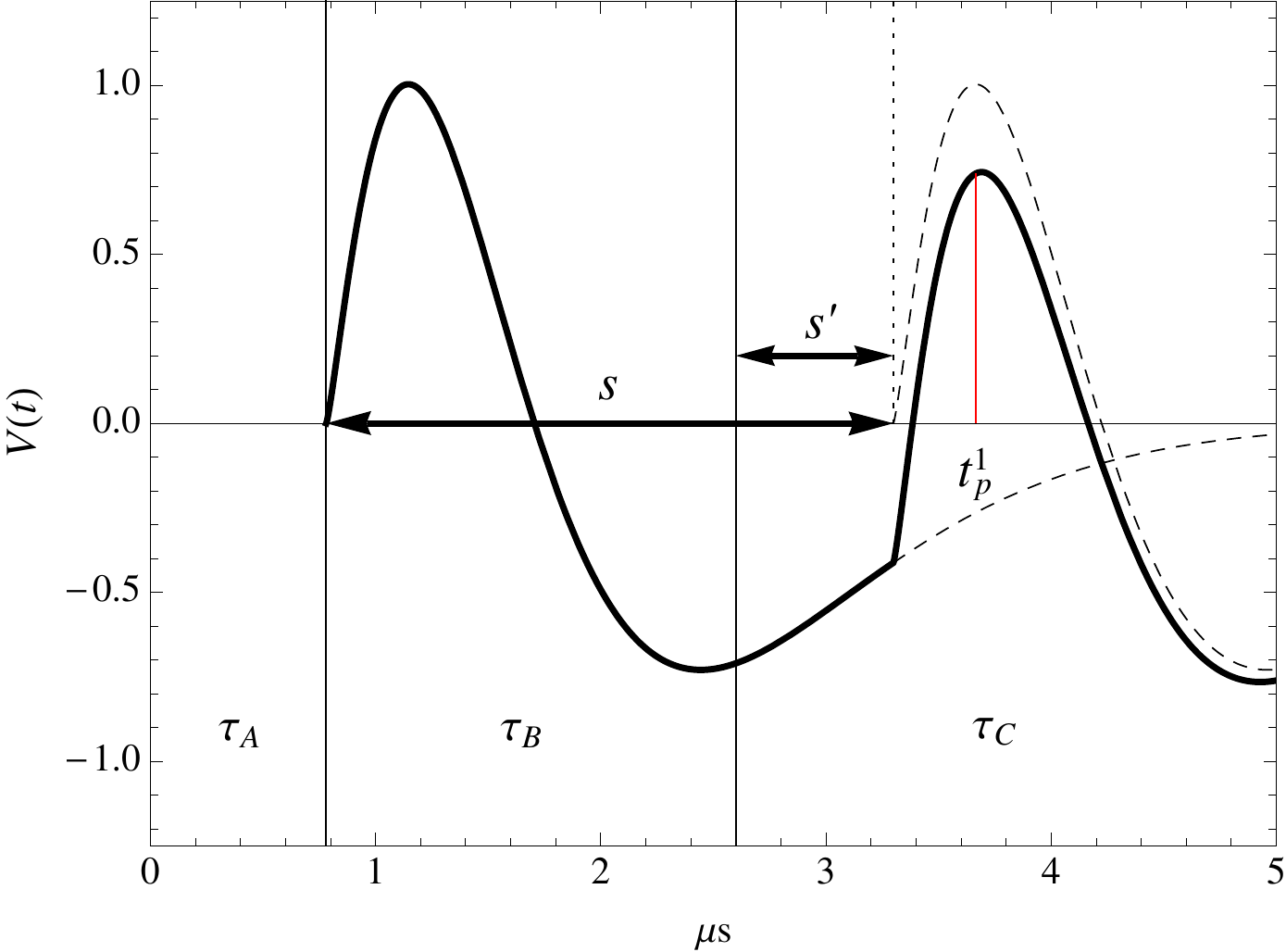}
\caption{Configuration used to calculate $\text{Pr}_{B+C}^{\langle1 \rangle}(\varepsilon | E_0,E_1)$, which assumes no zeroth event.  $t^1_p \approx t^0_p + s_1$ for the tail pulse.  $s^{\prime}$ is distributed according to equation \eqref{TailIntervalDist}.  }
\label{fig:BCPulseIntervals}
\end{figure}

%%%
\subsubsection{Second tail effect, $Q_{ \langle a,b,c \rangle}(\varepsilon)$  }

The second effect models the case when events in B pileup and create a large negative peak in C.  It is only present at orders $\geq 2$ since it requires events in B and C.  We approximate the variations due to $b$ random events in B using the peak spectrum terms calculated for order $b-1$ (there is no zeroth event in B).   Figure \ref{fig:BCPulseIntervals} depicts the lowest order B+C  configuration, used to calculate $\text{Pr}_{B+C}^{ \langle c=1 \rangle }$ by varying $s_1^{\prime}$.  Equation \eqref{eq:1stOrderTailEnergy} is used for the recorded energy, with the separation given as $s = s^{\prime} + \tau_B$, with $s^{\prime}$ distributed according to equation \eqref{TailIntervalDist}.   The probability of the pulse height lowering into $[ \varepsilon,\ \varepsilon + \Delta \varepsilon]$, due just to an event in B, is
\begin{equation}
\text{Pr}_{B+C}^{ \langle c \rangle }{(\varepsilon \ |\ E_{b-1} , E_{c} )} =  \int \limits_{s(\varepsilon)}^{s(\varepsilon + \Delta\varepsilon)}   f_{s|c}(s_1^{\prime})    ds^{\prime} 
\end{equation}
for $0 \leq s \leq \tau_C$. 

Finally, the additional event in C is an event from the input spectrum, shifted down due to tails from A, but also shifted up due to pulses in C.  In other words, it is approximately distributed as $q_{\langle a,0,c \rangle}(\varepsilon)$.  Thus we use these two terms with the kernel above to calculate the second component
%Equation: Q_abc
\begin{framed}
For $a \geq 0, \ b > 0, \ c > 0$,
\begin{dmath}\label{eq:TailCorrQabc}
Q_{\langle a,b,c \rangle}(\varepsilon) = \int \limits_{0}^{\infty} \int \limits_{0}^{\infty}  \text{Pr}_{B+C}^{ \langle c \rangle }(\varepsilon \ |\ E^{\prime}, E^{\prime\prime}) \times  p_{\langle b-1 \rangle}(E^{\prime})q_{\langle a,0,c \rangle}(E^{\prime\prime}) dE^{\prime} dE^{\prime\prime}
\end{dmath}
\end{framed}

This spectral component clearly depends on the total positive signal from both A and B.  Thus we have an approximation for the highest order states using the succession of terms, $Q_{\langle 0,0,1 \rangle} \rightarrow Q_{\langle 0,0,c \rangle} \rightarrow Q_{\langle a,0,c \rangle} \rightarrow Q_{\langle a,b,c \rangle}$.

%%%%%%%%
%	Probability of states
%%%%%%%%
\section{Probability of pileup events}\label{StateIndependence}

Since the time intervals A, B, and C do not overlap, the number of events in each interval is independent of the other two.  Furthermore, they are Poisson distributed random variables.  We assume $\lambda$ is constant throughout the detection of the input spectrum $S(E)$. The probability of each state is then
\begin{dgroup*}
\begin{dmath*} \text{Pr}( \text{state}\ |\ \text{rate}=\lambda  ) \end{dmath*}
\begin{dmath*}=\text{Pr}(a \hiderel{\in} A \text{ and } b  \hiderel{\in} B  \text{ and } c  \hiderel{\in} C\ |\ \lambda)\end{dmath*}
\begin{dmath*}= \frac{ (\lambda\tau_{A})^{a}(\lambda\tau_{B})^{b}(\lambda\tau_{C})^{c}   }{ a!*b!*c! } e^{-\lambda(\tau_{A}+\tau_{B}+\tau_{C}) }\end{dmath*}
\begin{dmath}\equiv \text{Pr}(\langle a,b,c \rangle | \lambda )\end{dmath}
\end{dgroup*}
The rate $\lambda$ is separately modeled so the condition can be ignored and the expression for state probability is
\begin{framed}
\begin{dmath}\label{eq:StateProb}
\text{Pr}( \langle a,b,c \rangle ) = \ \frac{ (\lambda\tau_{A})^{a}(\lambda\tau_{B})^{b}(\lambda\tau_{C})^{c}   }{ a!*b!*c! } e^{-\lambda \Delta \tau }  \condition[\ ]{    for $\tau_{A}+\tau_{B}+\tau_{C} = \Delta \tau$ }
\end{dmath}
\end{framed}
This gives the probability of having non-overlapping pulse intervals of width $\Delta \tau$ with the state $\langle a,b,c \rangle$.  These states are \emph{distinct}, meaning a single pulse window $\Delta \tau$ can only be `in' or `described by' a single state. Such propositions define a \textit{state space}, which unifies the total \emph{time} process with the total set of configurations.  In this scheme the total exposure time is decomposed into pulse windows of width $\Delta \tau$, and since every window has a state, the temporal composition is equivalent to a superposition of independent states.  The fraction of windows with order $k$ pileup is
\begin{dmath}\label{kthOrderProb}
\frac{ (\lambda \Delta \tau)^{k}  }{k! } e^{-\lambda \Delta \tau } = \sum_{i=0}^{k} \sum_{j=0}^{k-i}   \text{Pr}( \langle i, k - (i+j), j \rangle )
\end{dmath}
where the right-hand-side is the sum over the various combinations with $a+b+c = k$.  In the next section it will become clear why each order is sub-divided.  The terms above are treated as expansion coefficients with the spectral components appended.  

The total time process is just the superposition of independent states of order $k$, and thus
\begin{equation}
\sum_{k=0}^{\infty}  \frac{ (\lambda \Delta \tau)^{k}  }{k! } e^{-\lambda \Delta \tau } =  \sum_{a,b,c}^{\infty}  \text{Pr}\langle a,b,c \rangle = 1
\end{equation}
For example, if there are $N$ events incident during a long exposure time, the average number of non-overlapping windows $\Delta\tau$ with configuration $\langle n_A,n_B,n_C \rangle$ is equal to $N*\text{Pr}{\langle  n_A,n_B,n_C \rangle}$.

This section has assumed that the pulse windows $\Delta \tau$ are non-overlapping.  In the next section, we consider why the total spectrum is not correctly reconstructed under this assumption.  The error results from the fact that tail events require additional recovery time extending past the initial pulse window.  We will develop a semi-empirical probability expression that models pulse extension due to tail events by overlapping two windows.

%If we add to this the assumption that, for any given pulse, the entire pulse window $(\tau_A + \tau + \tau_C\equiv \Delta\tau)$ does overlap another pulse window, then states $\langle a,b,c \rangle$ are independent and we can describe the \textit{total state} of the system using a linear combination.  In the language of probability $\langle a,b,c \rangle$ are `independent events', and we can refer to $\langle a,b,c \rangle$ as the `event' of $(a+b+c)^{th}$ order pileup in a pulse window $\Delta\tau$. 

%%%%%%%%%
%
%  EXPANSION
%
%%%%%%%%%
\section{Full correction expansion}\label{FullExpansion}
The final correction is a combination of the $P$ and $Q$ terms and their associated state probabilities.  We first describe the total observed spectrum (i.e., the total process), as a superposition of independent (non-overlapping) pulse measurements.  We then devise a correction term based on the fact that measurement states overlap, due to the possibility of recording a tail event.

%%%%%%
% Subsection:
%	Independent states
%
\subsection{Independent states}
Adopting the assumptions that pileup states are independent, the total spectrum can be written as a superposition of the peak and tail components derived given in sections \ref{nthOrderPeakSpecDerive} and \ref{TotalTailSpectrum}.  For $k^{th}$ order pileup, $a+b+c=k$, and the $k$ events can be combined into A, B, C to give the set of possible pileup states $\Omega_k = \{\ \langle a,b,c \rangle\ :\ a+b+c=k \}$.  The total number of states per order $k$ is $\frac{(k+1)(k+2)}{2}$.

The $k^{th}$ order spectrum is an expansion into the spectral components with the state probability as coefficients.  However, since there are $k+1$ events and a maximum of two can be recorded (one peak, one tail), each $p,q$ term is multiplied by $\frac{1}{k+1}$.  Now the $k^{th}$ order term is written
\begin{dgroup*}
\begin{dmath}
f^{ (k)  }(\varepsilon)=\frac{1}{k+1}\sum_{i=0}^{k} \sum_{j=0}^{k-i} \text{Pr}  \langle i,  \Delta^k_{ij} , j  \rangle \times \Bigl\{  p_{i}(\varepsilon) + q_{i, \Delta^k_{ij} , j }(\varepsilon)  \Bigr\}
\end{dmath}
\begin{dmath}
=\frac{1}{k+1}\left\{ f_p^{ (k)  }(\varepsilon) + f_q^{ (k)  }(\varepsilon) \right\}
\end{dmath}
\end{dgroup*}
where 
\begin{align*}
\Delta^k_{ij} \equiv&\ k-(i+j)\\
p_{\langle 0 \rangle}(\varepsilon) =&\ S(\varepsilon) \\
q_{\langle a,b,0 \rangle}(\varepsilon) =&\ 0
\end{align*}
and the $p$, $q$ terms have been collected:
\begin{align}
 f_p^{ (k)  }(\varepsilon) =& \sum_{i=0}^{k} \sum_{j=0}^{k-i} \text{Pr}  \langle i,  \Delta^k_{ij} ,  j \rangle  p_{i}(\varepsilon) \\ 
 f_q^{ (k)  }(\varepsilon) =& \sum_{i=0}^{k} \sum_{j=0}^{k-i} \text{Pr}  \langle i,  \Delta^k_{ij} ,  j \rangle  q_{i,  \Delta^k_{ij} ,j}(\varepsilon)
\end{align}

The total spectrum (PDF), assuming independent pulse states, up to order $n$, is 
\begin{align}
f(\varepsilon) =& \sum \limits_{k=0}^n  f^{ (k)  }(\varepsilon)\\
=& \sum \limits_{k=0}^n \frac{1}{k+1}\left\{ f_p^{ ( k ) }(\varepsilon) + f_q^{ ( k ) }(\varepsilon) \right\}
\end{align}
The approximation order $n$ is the value at which $\frac{(\lambda \Delta \tau)^{n}}{n!} e^{-\lambda \Delta \tau}$ becomes negligible. 
%
%  FIGURE 11:  
%	Pulse overlap
\begin{figure}[h!]
\centering
\includegraphics[width=0.9\linewidth]{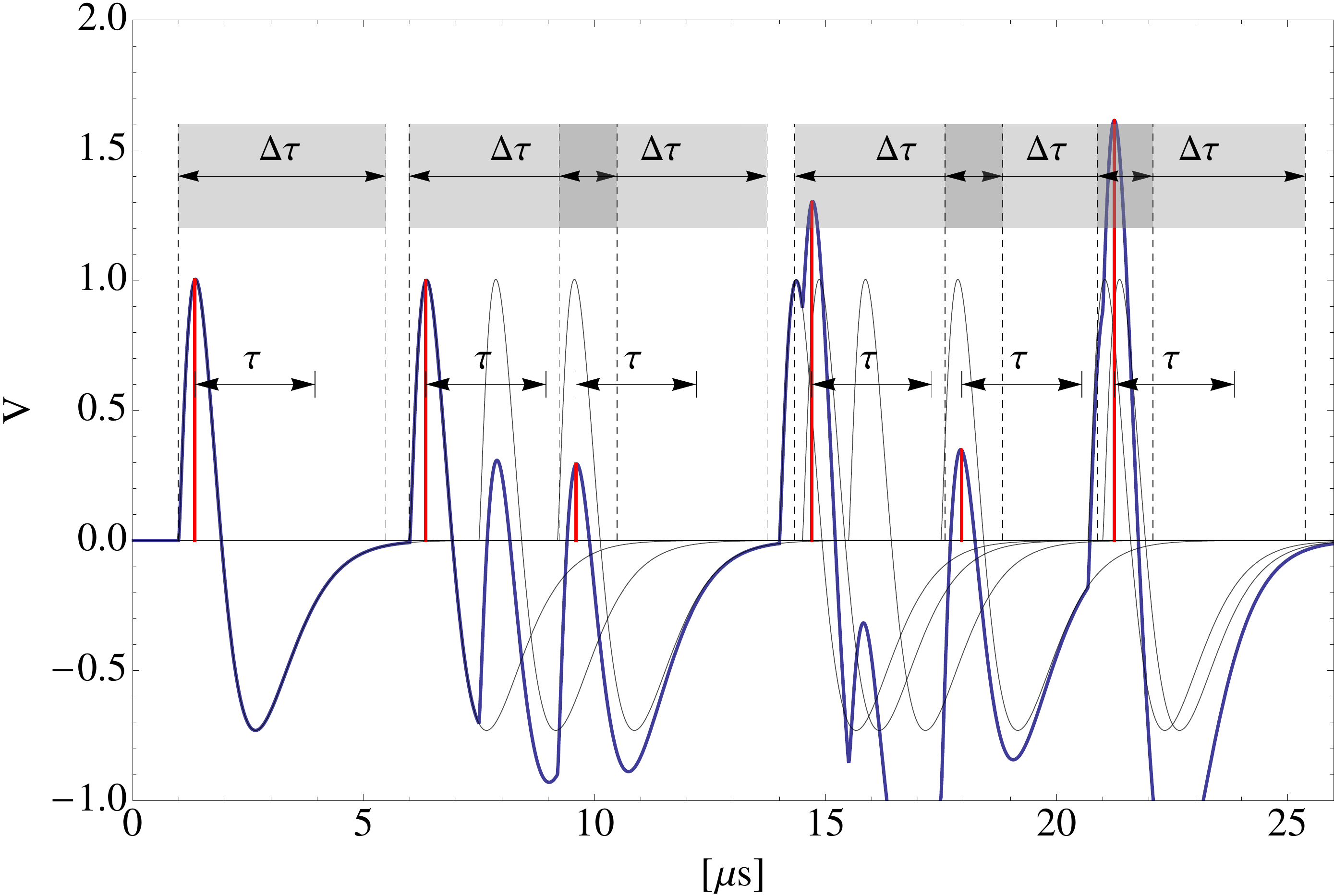}
\caption{ Sample output $V(t)$ showing how exposure time is partitioned into pulse states, with measured energies distributed as $p_a (\varepsilon)$ \& $q_{abc} (\varepsilon)$.  Subsequent pulse heights within $\Delta \tau$ of a tail measurements are distributed as $q_{abc} (\varepsilon)$.  A combination of independent states must be adjusted for the fact of overlapping regions. }
\label{fig:OverlapPulse}
\end{figure}

%
%  FIGURE:  
%	Independence vs dependence
%\begin{figure}[h!]
%\centering
%\includegraphics[width=0.95\linewidth]{Gauss2p2MeV_Correction_385kHz.eps}
%\caption{Comparison of the spectrum assuming non-overalapping pulse states (\emph{independent}) vs. the spectrum corrected for overlap (\emph{dependent}).  The input spectrum is a single gaussian-shaped line. }
%\label{fig:IndepVsDepCompareGauss}
%\end{figure}

%
%  FIGURE 12:  
%	Independence vs dependence
\begin{figure}[h!]
\centering
\includegraphics[width=0.95\linewidth]{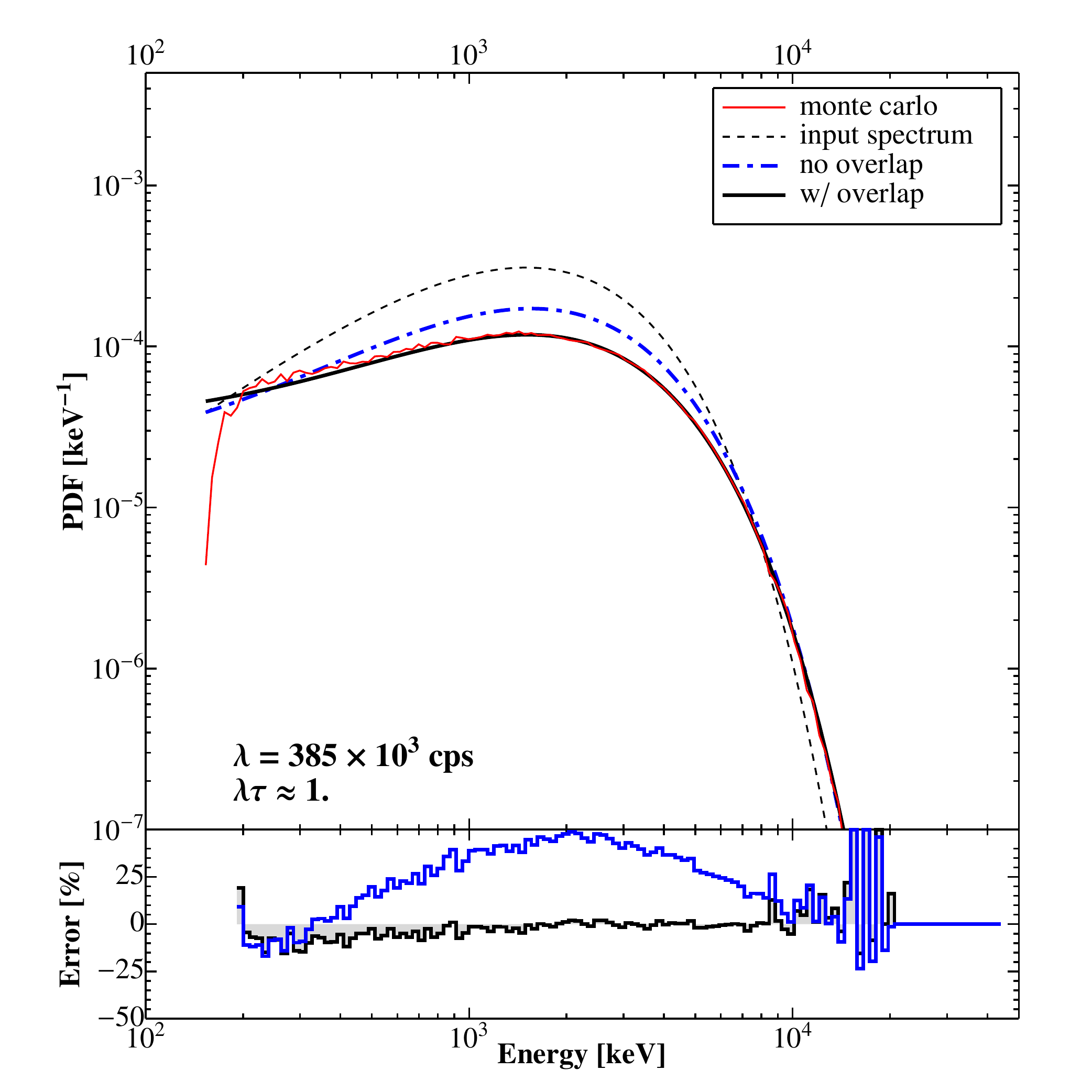}
\caption{Comparison with Monte Carlo simulation of the model spectrum assuming non-overalapping pulse states (BLUE, dashed) vs. the spectrum corrected for overlap (BLACK), for GBM.  The spectrum is a cutoff power-law, $S(E) \sim E^{1.5} \exp(-E / 1\text{ MeV})$.  Errors are calculated relative to the Monte Carlo:  (model - MC) / MC }
\label{fig:IndepVsDepCompareCutPL}
\end{figure}

%%%%%%
% Subsection:
%	Dependent states
%
\subsection{Overlapping pulses (dependent states)}\label{Overlap}
This expansion is accurate at low rates, when the probability of having two adjacent pileup states is small.  But we must account for the fact that a tail measurement causes a new pulse interval to `begin' before the end of the first.  A theoretically correct construction from the partitioned light curve (i.e., the total process) becomes rather complex due to this fact.  Each state with a tail count, by definition, also contains the `beginning' of the next pulse interval.  Events occurring within $\Delta \tau$ of the first tail count must also be modeled.  In terms of the state spectral components $p$ and $q$, the total measurement from this joined-pulse interval contains a peak count, a tail count, and a possible third tail count (from the adjoined pulse). 

One way to model this is to consider all possible combinations of states.  If $\Omega$ is the set of possible states, then we would have to partition the process in terms of the product states $\Omega \otimes \Omega$.  At very high rates it may even be necessary to consider $\Omega \otimes \Omega \otimes \Omega$, since additional pulse can occur with $\Delta \tau$ of the third tail measurement.  Due to the obvious complications, we have developed a semi-empirical approximation which adjust the weighting of peak and tail components.

In the approximation two overlapping intervals can be represented as dependent random events $M$ and $M^{\prime}$, whose probability is generally constructed as
\begin{equation}
\text{Pr}(M \cup M^{\prime}) = \text{Pr}(M) + \text{Pr}(M^{\prime}) - \text{Pr}(M \cap M^{\prime})
\end{equation}
In general the adjoined state, which we may identify as  $M^{\prime} = \langle a^{\prime},b^{\prime},c^{\prime} \rangle$ can be begin at any random time in the interval $\tau_C$, if M contains a tail pulse.  Thus $M = \langle a,b,c \rangle$ with $c \geq 1$ (and any $a,b$).  Figure \ref{fig:OverlapPulse} shows several pulse windows with tail measurements.
For a given state $\langle a,b,c \rangle$ with $c \geq 1$, the counts $c$ constitute the peak of a subsequent pileup event, denoted by the symbol $\langle c-1,b^{\prime},c^{\prime} \rangle$.  Then we approximate the probability $\text{Pr}(M \cap M^{\prime})$ as the probability of having $c$ counts in the tail of a state $M$ \textbf{and} $c-1$ counts in the peak of a state $M^{\prime}$  (without regard to the $a,b$ or $b^{\prime},c^{\prime}$)
%
% EQUATION:    Overlap states
%
\begin{dgroup*}
\begin{dmath}
\text{Pr}\biggl(\langle abc \rangle \cap \langle (c-1) b^{\prime} c^{\prime} \rangle\biggr)   \approx  \frac{ ( \lambda \tau_C)^{c} e^{-\lambda \tau_{C} } }{c!} \times \frac{ (\lambda \tau_A)^{c-1} e^{-\lambda \tau_{A} }}{ (c-1)! } 
\end{dmath}
\begin{dmath}
\equiv \text{Pr}(c, a^{\prime}) \delta_{ c-1, a^{\prime}}
\end{dmath}
\end{dgroup*}%

%
% EQUATION:    Full correction
%

The corrected peak contribution is reduced by one count per configuration of this tail-peak overlap, so we have a single term subtracted at each order, for $k>0$, from $f^{(k)}_p$:
\begin{dgroup*}
\begin{dmath*}
f^{(k)}_{p,Corr}(\varepsilon) =
\end{dmath*}
\begin{dmath} \label{eq:fpCorrection}
\  \frac{1}{k+1} \Bigl\{ f_p^{(k) }(\varepsilon) - \text{Pr}(k+1,k) p_k(\varepsilon) \Bigr\}
\end{dmath} 
\end{dgroup*}
with
\begin{equation}
 \text{Pr}(k+1, k) = \frac{ (\lambda \tau_C)^{k+1} e^{-\lambda \tau_{C} } }{(k+1)!} \times \frac{ (\lambda \tau_A)^{k} e^{-\lambda \tau_{A} }}{ k! }
\end{equation}
For $k=0$ there is no tail count, thus no need to worry about carry-over effects. Therefore we can say
\begin{equation*}
f^{(k=0)}_{p,Corr}(\varepsilon) = f^{(0)}_p(\varepsilon) = e^{-\lambda \Delta \tau} S(\varepsilon)
\end{equation*}
which is the weighted zeroth order term giving the fraction of counts without peak pileup or a trailing deadtime / tail event.

The tail correction adds weight to the contribution of $f^{(k)}_q$ in the total spectrum, since the adjoined pulse has its own tail region.  The energy of these counts has a similar distribution to tail counts of the initial pulse.  Therefore we use the tail terms $q$, but adjust the associated probability.  We model it as the complement of the overlap probability all  configurations; i.e., $1 -  \sum_{k=0}^n \text{Pr}(k+1, k)$. The corrected tail contribution is:

%The reason for this second summation is that the adjoined pulse window, sticking with the A-B-C partition scheme, introduces two additional degrees of freedom, $b^{\prime}$ and $c^{\prime}$.  And because these are tail events, the recorded spectrum of pulse heights will be very similar to the total tail spectrum already computed for independent windows, $f^{(k)}_q$. All that remains is a proper weighting.  We find that, if $k^{\prime}$ represents the pileup order in the overlap with probability $\text{Pr}(k^{\prime}+1, k^{\prime})$, the added probability is approximately equal to the independent-state probability, condo
\begin{dgroup*}
\begin{dmath*}
f^{(k)}_{q,Corr}(\varepsilon)=
\end{dmath*}
\begin{dmath}
\frac{1}{k+1} \Bigl\{ \Bigl[1 -  \sum \limits_{k^{\prime}=0}^{\infty} \text{Pr}(k^{\prime}+1, k^{\prime}) \Bigr] f^{(k)}_{q} + f^{(k)}_{q} \Bigr\}
\end{dmath}
\begin{dmath}
= \frac{1}{k+1} \Bigl[2 -  \sum \limits_{k^{\prime}=0}^{\infty} \text{Pr}(k^{\prime}+1, k^{\prime}) \Bigr] f^{(k)}_{q}
\end{dmath}
\end{dgroup*}
Again for $k=0$ there are no tail counts, so $f^{(0)}_{q,Corr}(\varepsilon)=f^{(0)}_{q}(\varepsilon)=0$.  Now the total measured spectrum is written
\begin{framed}
\begin{dgroup*}
\begin{dmath*}
f(\varepsilon)= f^{(0)}_{p}(\varepsilon)  + \sum \limits_{k=1}^n  \frac{1}{k+1} \times
\end{dmath*}
\begin{dmath}\label{eq:FullCorrectedSpectrum}
 \biggl\{ f_p^{(k) }(\varepsilon) - \text{Pr}(k+1,k) p_k(\varepsilon) +   \Bigl[2 -  \sum \limits_{k^{\prime}=0}^{\infty} \text{Pr}(k^{\prime}+1, k^{\prime}) \Bigr] f^{(k)}_{q} (\varepsilon) \biggr\}
\end{dmath}
\end{dgroup*}
\end{framed}
or simply
\begin{equation}
f(\varepsilon)=  \sum \limits_{k=0}^n \Bigl\{ f^{(k)}_{p,Corr}(\varepsilon) + f^{(k)}_{q,Corr}(\varepsilon) \Bigr\}
\end{equation}

This method is evaluated by comparing computed spectra with Monte Carlo simulations.  Figure \ref{fig:IndepVsDepCompareCutPL} demonstrates a spectrum computed under the assumption of independent pulse states, versus the overlap correction.  The Monte Carlo is described with more comparisons in section \ref{MonteCarloSection}.  Losses due to pileup and subtraction effects can be summarized as follows.  If the true detection rate is $\lambda_{0}$, the predicted counting rate is
\begin{align}
\lambda_{Rec} &=  \lambda_{0} \int f(\varepsilon) d\varepsilon \\
& = \lambda_{0}  \sum \limits_{k=0}^n \int f^{ (k) }(\varepsilon) d\varepsilon 
\end{align}
The relationship between $\lambda_{Rec}$ and $\lambda_{0}$ is discussed further in section \ref{PileupLosses}.

\section{Numerical evaluation}\label{Evaluation}
In this section we apply the model to GBM and compare it with Monte Carlo simulations.  The first step is numerical evaluation of the probability likelihoods, which are used to derive pileup spectral components.

% Section: GBM Peak Time
%
\subsection{Peak time}\label{GBMPeakTime}
The single-pulse peak time is given by $f^{\prime}(t^0_p) = 0$.  Recall that the first-order peak time, defined in general by equation \eqref{eq:1stOrderPeakGeneral}, cannot be found in closed form for the true pulse shape $f(t)$.  Instead we make two approximations of the function $t^1_p(s_1,v_0,v_1)$ for region A and region C, and join them smoothly in region B.  For peak pileup (A), we use an empirical expression with two constant parameters $\Lambda_s, \Lambda_v$.  Events are sampled on a $(s_1,v_0,v_1)$ grid, and PHA measurement is simulated with the same routines used in the Monte Carlo.  Resulting peak times are fit using a non-linear least-squares technique and the functional form below: 

\begin{equation}
t^1_p(s_1,v_0,v_1) =\bar{t^0_p}\ +\ F(  e^{ \Lambda_s s_1^2 } - 1  )  e^{ - \Lambda_v ( \frac{ v_0 - v_1 }{ v_0 + v_1 } )^2 }
\label{eq:t1p}
\end{equation}
where $\bar{t^0_p}$ is the weighted average of the zeroth-order peak time of each pulse,
\begin{equation}
\bar{t^0_p} = \frac{ v_0  t^0_p + v_1 (t^0_p + s_1) } { v_0 + v_1 } = t^0_p + \frac{v_1}{v_0 + v_1} s_1
\end{equation}
and $F=1\ \mu s$.  The second term adjusts $\bar{t^0_p}$. The best-fit parameters are $\Lambda_v = 2.09, \Lambda_s = 0.40$, for $s_1$ in $\mu s$.  Of course these parameters, and the functional form itself, are dependent on the specific pulse shape.  The ones given here correspond to the best-fit GBM pulse shape from section \ref{GBMPulseShape}.

In region C we use the approximation $t^1_p \approx t^0_p + s_1$.  For region B these two expressions are smoothly joined using a logistic function about $s_1 = 1 \ \mu s$, which is approximately the positive pulse width.  The joining scale is an arbitrary parameter, which we set to $\Lambda_{0} = 30 [\mu s]^{-1}$.  Thus the full peak time expression is:
\begin{dmath}
t^1_p(s_1,v_0,v_1) = \Bigl[ \bar{t^0_p}\ +\ (  e^{ \Lambda_s s_1^2 } - 1 )  e^{ - \Lambda_v ( \frac{ v_0 - v_1 }{ v_0 + v_1 } )^2 }  \Bigr] \times \Bigl(1 - g(s_1) \Bigr) +  \Bigl[ t^0_p + s_1  \Bigr] \times g(s_1)
\label{eq:TotalPeakTime}
\end{dmath}
where $g(s_1) = 1 / [ 1 + \exp( -\Lambda_{0} (s_1 - 1))]$.  

Finally, in the case of peak-pileup \emph{only} (i.e., $v_0, v_1$ in A), under certain conditions the initial pulse is accurately distinguished from the summed signal.  This results in a piecewise step in the peak-time formula such that $t^1_p \rightarrow t^0_p$, observable as discontinuity in region A of figure \ref{fig:MeasuredEnergy1}.  This generally occurs when $v_0 > v_1$ and the separations are sufficiently large.  An inspection of the simulated peak-time data reveals that the precise condition is a complicated function of all three variables, however its strongest dependence is on $s_1$.  By manual estimation we have determined the following additional criteria for the peak-pileup time:  \textbf{if} $t^1_p > t^0_p * \exp(1.7 s_1^2)$, where $t^1_p$ is from equation \eqref{eq:t1p}, \textbf{or} $s_1 > \tau_A$, \textbf{then} set $t^1_p \rightarrow t^0_p$.  This is approximates the effect of digital buffering described in section \ref{GBMPulseShape}, and is only for the case of peak-pileup.

% Section: Discrete Calc.
%
\subsection{Calculation of $\text{Pr}_{\rm peak}^{ \langle n \rangle }, \text{Pr}_{A+C}^{ \langle n \rangle }, \text{Pr}_{B+C}^{ \langle n \rangle }$}\label{GBMProbTensors}
Because we are using the more accurate pulse shape, the conditional probability kernels cannot be calculated in closed form.  We calculate them numerically by first first defining a discrete set of channel energies in which probabilities are calculated, $\textbf{E} = \{E_i\}$.  Since pileup events have been detected by the instrument it's sufficient to use the channel definitions of actual data.

Defining a discrete time step $\Delta s$, integrals like 
\begin{equation}
\text{Pr}(\varepsilon | E, E^{\prime} ) = \int f_{s|n}(s^{\prime}) ds^{\prime}
\end{equation}
are evaluated discretely over the finite $\Delta s$ sample ($\textbf{s} = \{ 0, \Delta s, 2\Delta s, \dotsc, l \Delta s \}$).  At each step of the integrand, the function $\varepsilon_1(s | E, E^{\prime} )$ is evaluated.  Then $\varepsilon_1$ corresponds to one of the channels in \textbf{E}, and an appropriate lookup algorithm returns its index (channel) $i$.  For $E, E^{\prime}$ also discretely sampled from \textbf{E} and corresponding to channels $j,k$, the conditional probability is a triply-indexed discrete object $\text{Pr}(\varepsilon | E, E^{\prime} ) \rightarrow \text{Pr}[i,j,k]$.  

This probability `array' is initialized to 0 for all $i,j,k$, and incremented by a value $\Delta P_l$ for each step in the RHS numerical integration, where $l$ specifies the integration step $s_l$ to $s_l + \Delta s$, and $i$ is the channel corresponding to $\varepsilon_1( s_l | \  \textbf{E}_j, \textbf{E}_k )$.  The interval distributions are integrable, so $\Delta P_l$ is exact for the $l^{th}$ time step:
\begin{align*}
\Delta P_l &= \int \limits_{s_l}^{s_{l+1}} \frac{n}{\tau^n}(\tau - s^{\prime})^{n-1} ds^{\prime}\\
&= \frac{ (\tau - s_l)^n - (\tau - s_{l+1})^{n} }{ \tau^n }
\end{align*}
where $\tau$ is the normalization interval (A, B, or C) and $s_l \in [0,\tau]$.
The calculation and storage of the $\text{Pr}(\varepsilon | E, E^{\prime} )$ likelihoods introduces some computational overhead, since we must have one per order in each interval. But since no information about the source spectrum or intensity is required, they can be calculated \textit{a priori} and stored.  A separate code computing spectrum corrections can read each data block as necessary.  In our implementation we use 128 energy channels and store each function in its own 3D data set in an HDF5 file \cite{hdf5}.  This library was chosen for its stability and ease-of-use in storing a set of multi-dimensional arrays and keyword parameters in a single file.  Using 4-byte floating point data, the total uncompressed requirement for all orders up to 5 is $(3*128^3)\times(4\text{ bytes})\times 5 = 125 \text{MB}$, which is well within the available memory in most environments.

%1048576
% Section:  MONTE CARLO
%
\subsection{ Monte Carlo comparison } \label{MonteCarloSection}
We have implemented a Monte Carlo simulation that includes both the arrival of random events and the discrete pulse-height measurements made by the instrument.  We simulate a sequence of exponentially distributed event times over an arbitrary exposure interval, during which the process intensity is constant.  Their energies are sampled from an input spectrum $S(E)$, which represents the recorded spectrum in the \emph{absence} of pileup distortion.  Our simulation includes the processing of signals by the GBM pulse-height electronics, but not physical interaction of gamma-rays in the detectors.  Because the detector response is neglected, the example input spectra shown in the next section are idealized, e.g., monochromatic gamma-rays would not generate a pure Gaussian input spectrum because, for some of the incident photons, only a portion of the energy is deposited in the crystal (Compton scattering or pair production), and the remaining energy might escape.  This information is summarized in the detector response information that converts an external gamma-ray spectrum into the input spectrum $S(E)$ (equation \eqref{eq:ResponseConvolution}), and is not a focus of this paper.

The analytical pileup model is calculated up to order 5.  For $k > 5$, an approximation is used by substituting fifth-order likelihoods in for higher orders in the iterative calculation.  At high orders this results in a slight overestimation for states having $a+b > 5 \gg c$.  Monte Carlo simulations are executed for 200,000 input events, at several different rates.  This simulation size is large enough to limit statistical fluctuations in the output spectra, allowing a more systematic comparison with the analytical model.

\begin{figure}[ht!]
\centering
\includegraphics[width=0.85\linewidth]{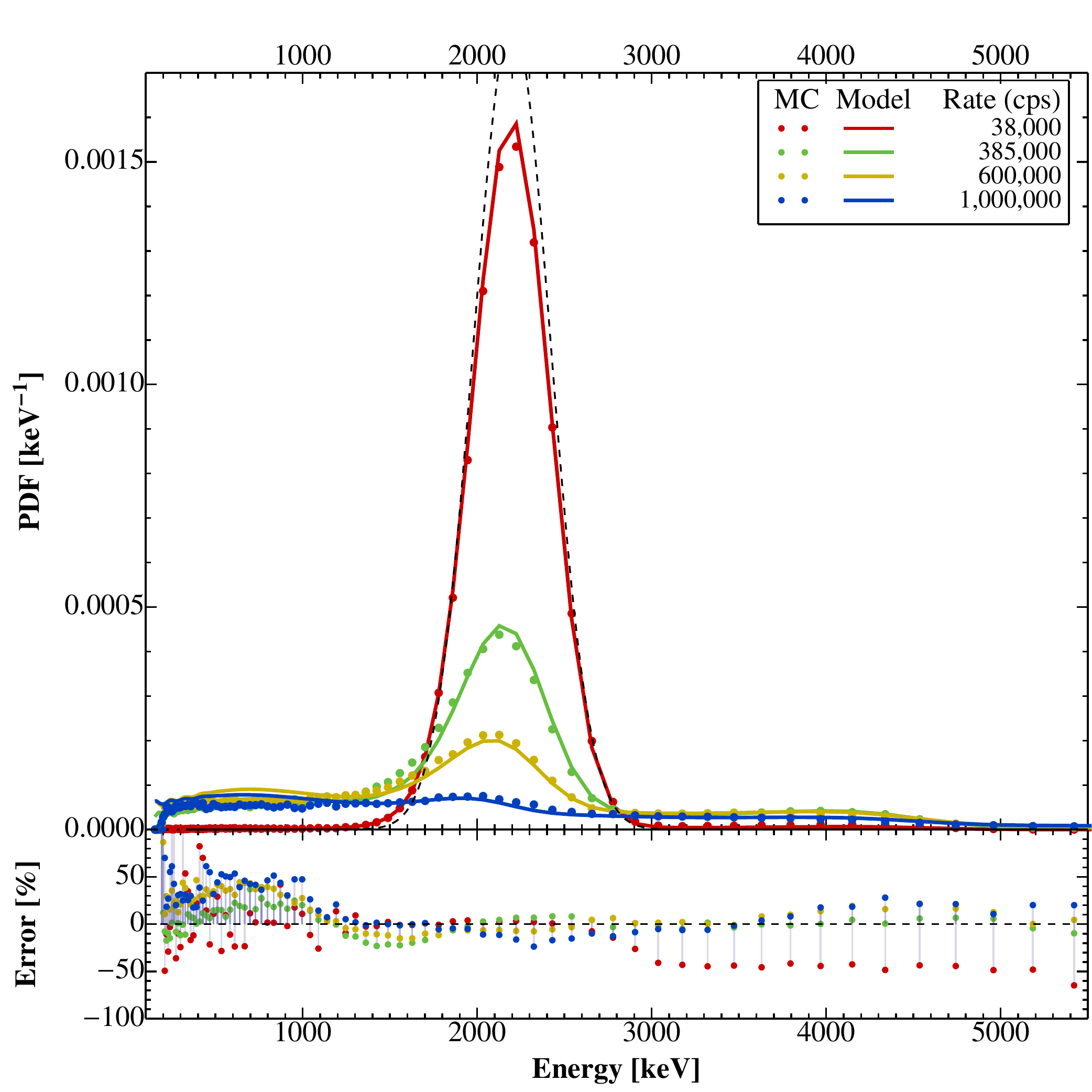}
\caption{Model vs. simulation comparison for several input rates. The input spectrum is a single Gaussian-shaped line at 2.2 MeV, and is shown by the black dashed line.  At high rates the line is completely distorted.  }
\label{fig:GaussCompared}
\end{figure}

An instructive case is that of a narrow line spectrum.  We simulate a pure Gaussian line with $\mu=2.2 MeV$ and $\sigma = 0.1 \mu$, and compare it with the model prediction (figure \ref{fig:GaussCompared} ).  At low rates the amount of spectral distortion is quite small, but as the rate increases the line becomes distorted.  Because $p_a(\varepsilon)$ and  $q_{abc}(\varepsilon)$ components are independent of the rate (only their expansion coefficients vary), a range of input rates can be tested for a given $S(E)$ without much difficulty.

\begin{figure}[ht!]
\centering
\includegraphics[width=0.85\linewidth]{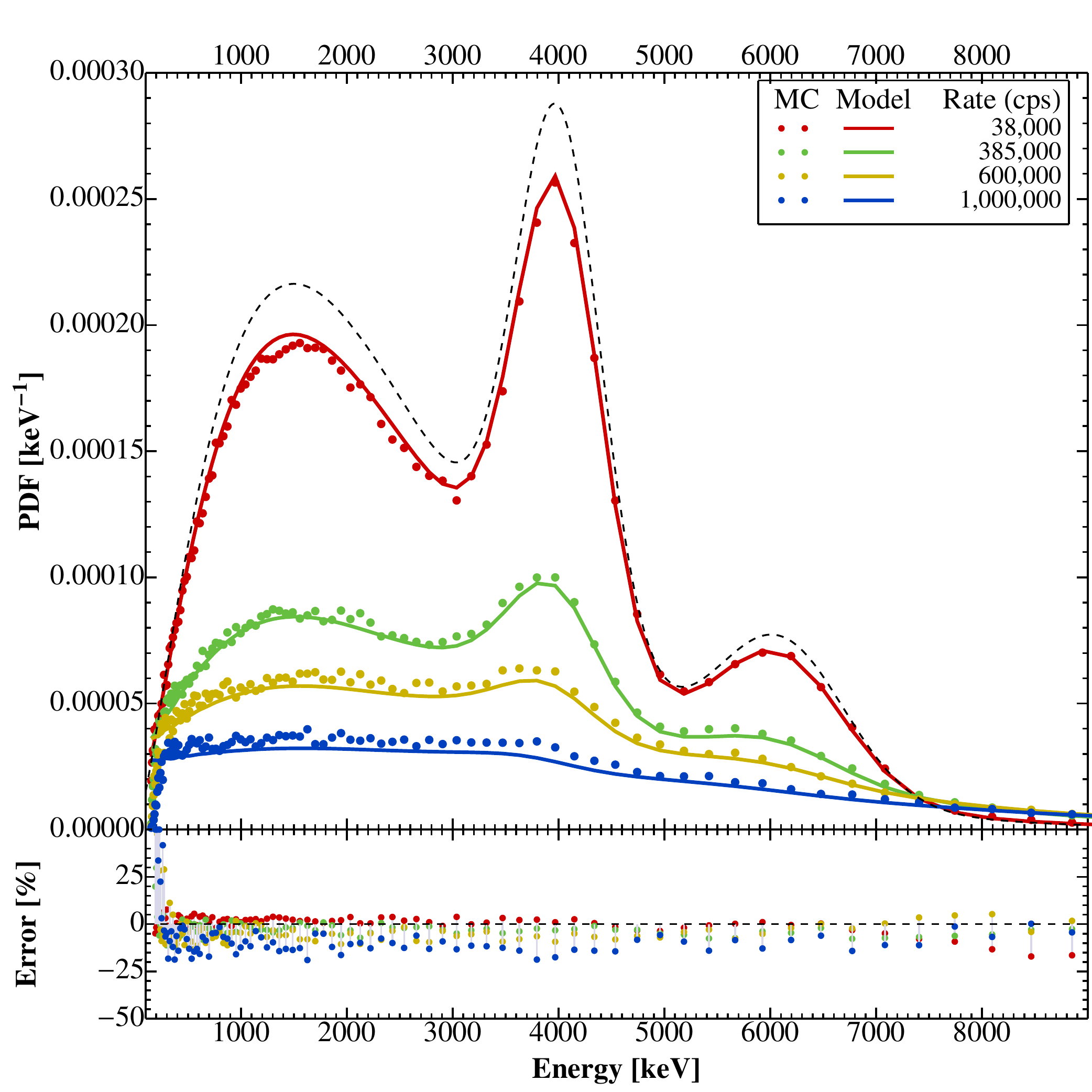}
\caption{A spectrum of two lines (4 and 6 MeV) on a cut-off power-law continuum.  The model becomes more accurate for spectra with a continuum, suggesting that energy raising and lowering associated with peak and tail modeling have canceling errors.   }
\label{fig:TwoLinesCutoff}
\end{figure}

Plot residuals (errors) are calculated relative to the Monte Carlo output.  If $n_i$ is the number of counts in the channel $i$ of the output MC spectrum, and $m_i$ is the analytical prediction (i.e., the model), residuals are calculated $r_i = (m_i - n_i) / n_i$ (times 100).  The model is reasonably accurate for the Gaussian spectrum, and improves for the other spectra shown.  For spectral shapes $S(E)$ dominated by narrow-band features, such as figure  \ref{fig:GaussCompared} where the line occupies about 1.6\% of the total voltage range, the model error tends to be higher in channels away from the line, where the only counts are due to high order pileup.  To the left of the line are primarily tail distorted counts, and to the right are mainly counts suffering from peak pileup.  However for broad-band models, such as the one in figure \ref{fig:TwoLinesCutoff}, the error is much smaller and more uniform.  The notable difference is that more channels have both peak \emph{and} tail pileup counts when the input spectrum is broad, suggesting that the peak and tail modeling assumptions have complementary errors, which tend to cancel out when combined.

\begin{figure}[ht!]
\centering
\includegraphics[width=0.95\linewidth]{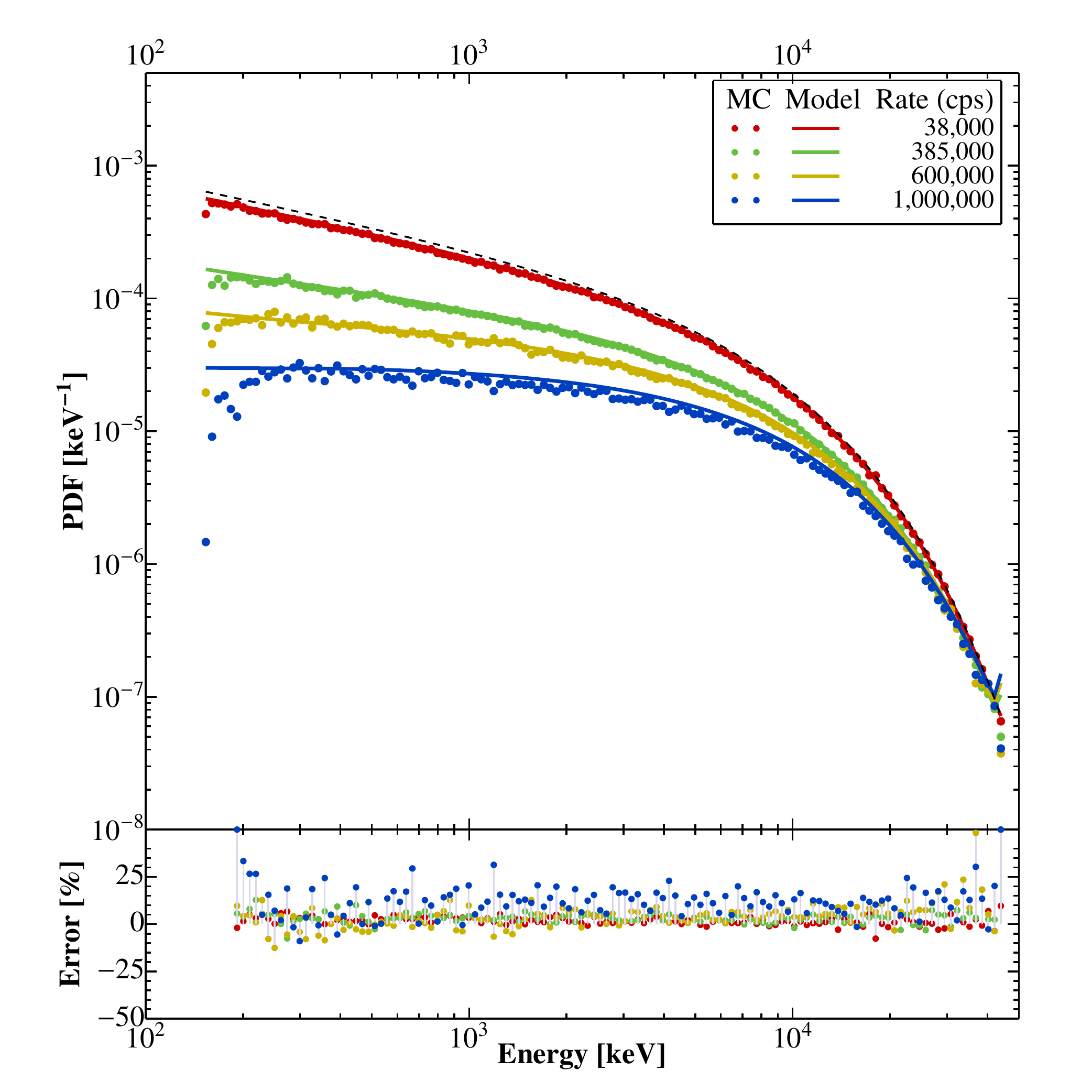}
\caption{A cut-off power law model, whose PDF is $\sim E^{-0.5}\exp( -E / 7 \text{MeV} )$, shown at several rates.  The spectrum is plotted on a log-log scale.}
\label{fig:RatesCompared7MeV}
\end{figure}

\begin{figure}[ht!]
\centering
\includegraphics[width=0.95\linewidth]{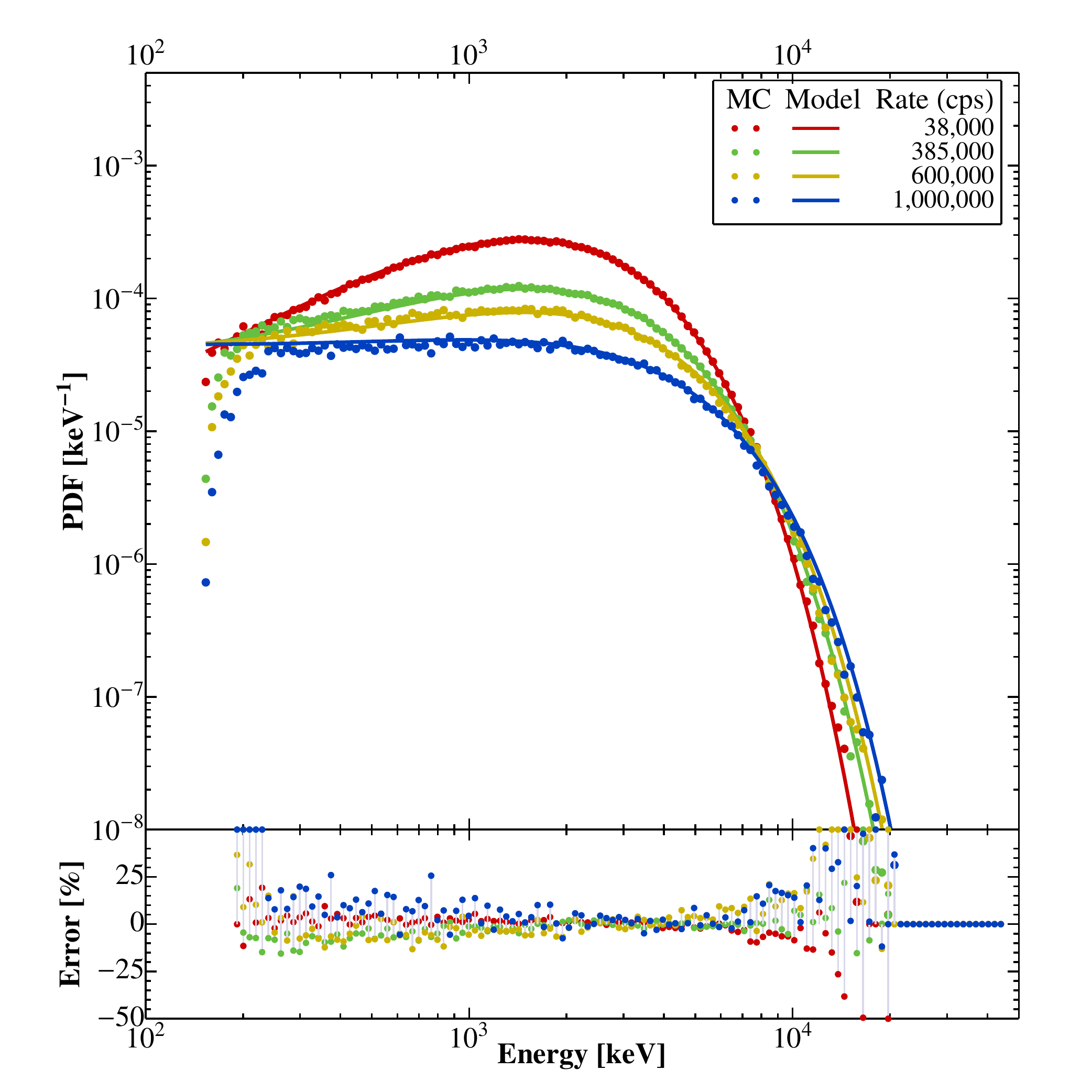}
\caption{A cut-off power law model, whose PDF is $\sim E^{1.5}\exp( -E / 1 \text{MeV} )$, shown at several rates on a log-log scale.}
\label{fig:RatesCompared1MeV}
\end{figure}

Figures \ref{fig:RatesCompared7MeV} and \ref{fig:RatesCompared1MeV} demonstrate the effects of pulse pileup when the zeroth order count spectrum is a smooth, cut-off power-law.  Low-energy spectral indices become flattened as lower energy pulses are shifted up through peak-pileup or removed through tail subtraction. The consequence is that pileup can significantly affect inferences about the power-law index, peak or cut-off energy parameters, and measures of spectral hardness unless corrected.  In general count spectra will become harder as the input rate increases. 

\begin{figure}[ht!]
\centering
\includegraphics[width=0.95\linewidth]{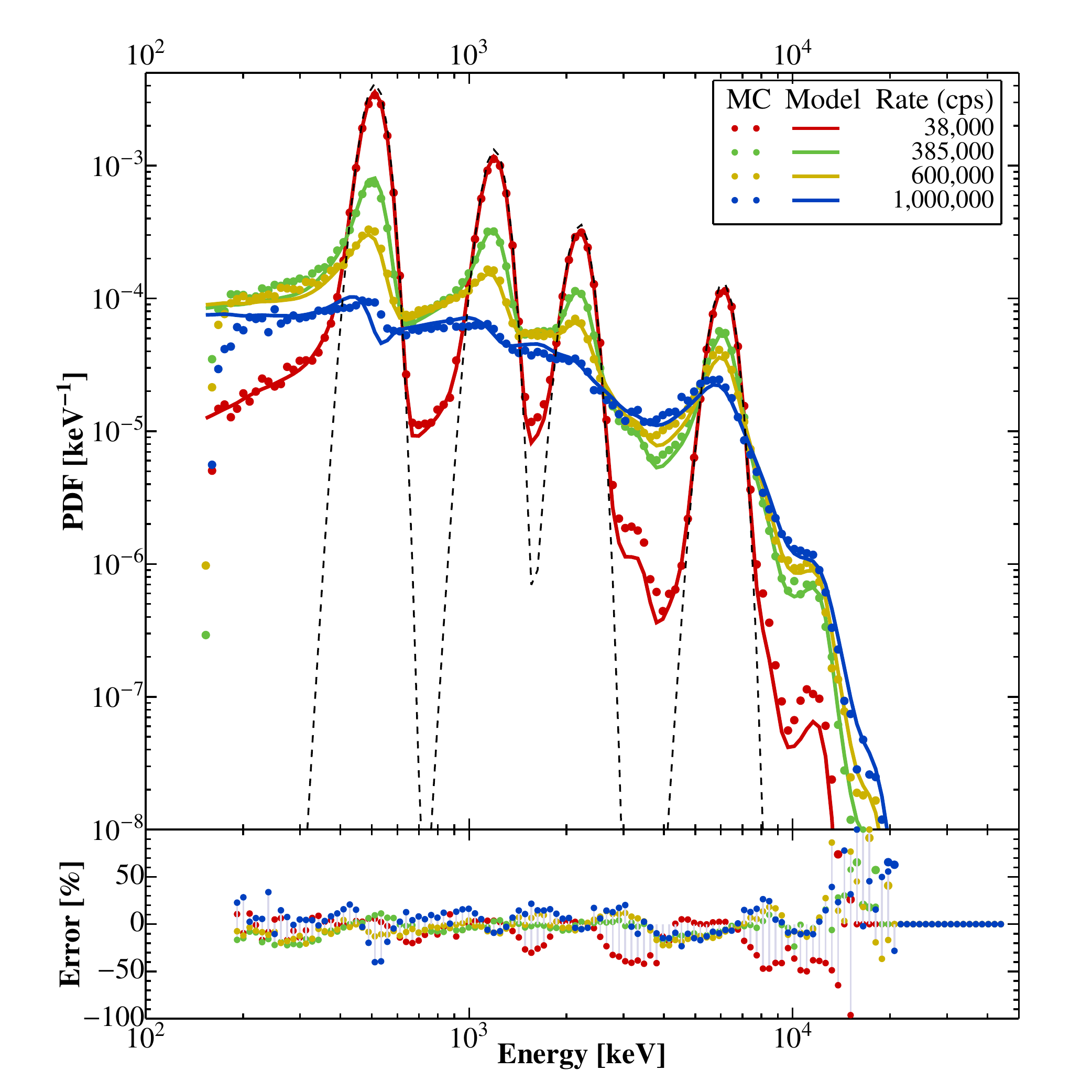}
\caption{A spectrum of four fictitious emission lines, with no continuum.   }
\label{fig:RatesCompared4Lines}
\end{figure}

Figure \ref{fig:RatesCompared4Lines} is a fictitious spectrum demonstrating four emission lines without a continuum.  Even without a continuum lines become distorted at high rates.  Additional false lines can appear from peak pileup of each line source.

Finally, the energy binning used to generate plots in this section are approximately those of the GBM BGO detectors.  However, the pileup process occurs in voltage space.  The model  

%
%  FIGURE 13:  
%	gauss

%
%  FIGURE 14:  
%	Two lines + continuum

%
%  FIGURE 15:  
%	Cutoff -0.5, 7 MeV

%
%  FIGURE 16:  
%	Cutoff  1.5, 7 MeV

%
%  FIGURE 17:  
%	Four Lines

%%%%%%%
% Section:  
%    Loss statistics
%
%
\section{Pileup losses}\label{PileupLosses}

As already mentioned in the opening sections pulse-pileup introduces counting losses which exceed the expectation from the conventional non-paralyzed deadtime correction.  By simple arguments this function is $\lambda_{Rec} = \lambda_{0} / (1 + \lambda_{0} \tau)$, and is algebraically invertible to give 
\begin{equation}\label{DeadtimeCorrection}
\lambda_0 = \lambda_{Rec} / (1 - \lambda_{Rec} \tau)
\end{equation}
At high rates it is insufficient both because of pileup, and because the derivation assumes $\lambda_{Rec} < 1 / \tau$. \cite{taguchi:2010}, \cite{WRLeo}

Pileup changes the picture for two reasons.  The first is that it randomly extends instrumental deadtime.  In GBM this occurs only when events are within $\tau_A$ of each preceding event, since the additional detections delay peak measurement.  The second is that the tail effect causes energy-dependent baseline subtraction losses, when a tail shifts small pulses below the recording threshold.  Such losses are more significant for `flat' or broadband energetic spectra because they entail a mixture of large and small pulses, and the smaller ones are more likely to be lost.  In GBM these additional losses are mitigated by the imposed 2.6 $\mu s$ deadtime, but at high rates they must be accounted for to fully correct the relation $\lambda_{Rec} (\lambda_0 )$.

The first effect (paralyzable deadtime) by itself can be predicted, to first order, using Poisson statistics and employing a root-finding algorithm.  The procedure is described in \cite{WRLeo}, and yields a numerical solution.  We do not explicitly use this treatment in our model.  Rather, because the model's predicted losses agree well with simulation, we conclude that the method of overlapping windows, equation \eqref{eq:FullCorrectedSpectrum}, is an effective proxy for paralyzable deadtime.  This seems reasonable since the subtracted overlap terms in equation \eqref{eq:fpCorrection} reduce the proportion of peak measurements, which is roughly the same effect as paralyzable deadtime.  An alternative interpretation is that severe paralyzable deadtime is improbable in GBM even at the high rates tested, due to the pulse shape and peak finding algorithm. At rates beyond $10^6$ cps, it is likely that the instrument becomes non-linear and our assumptions would not hold. 

\begin{figure}[h!]
\centering
\includegraphics[width=0.95\linewidth]{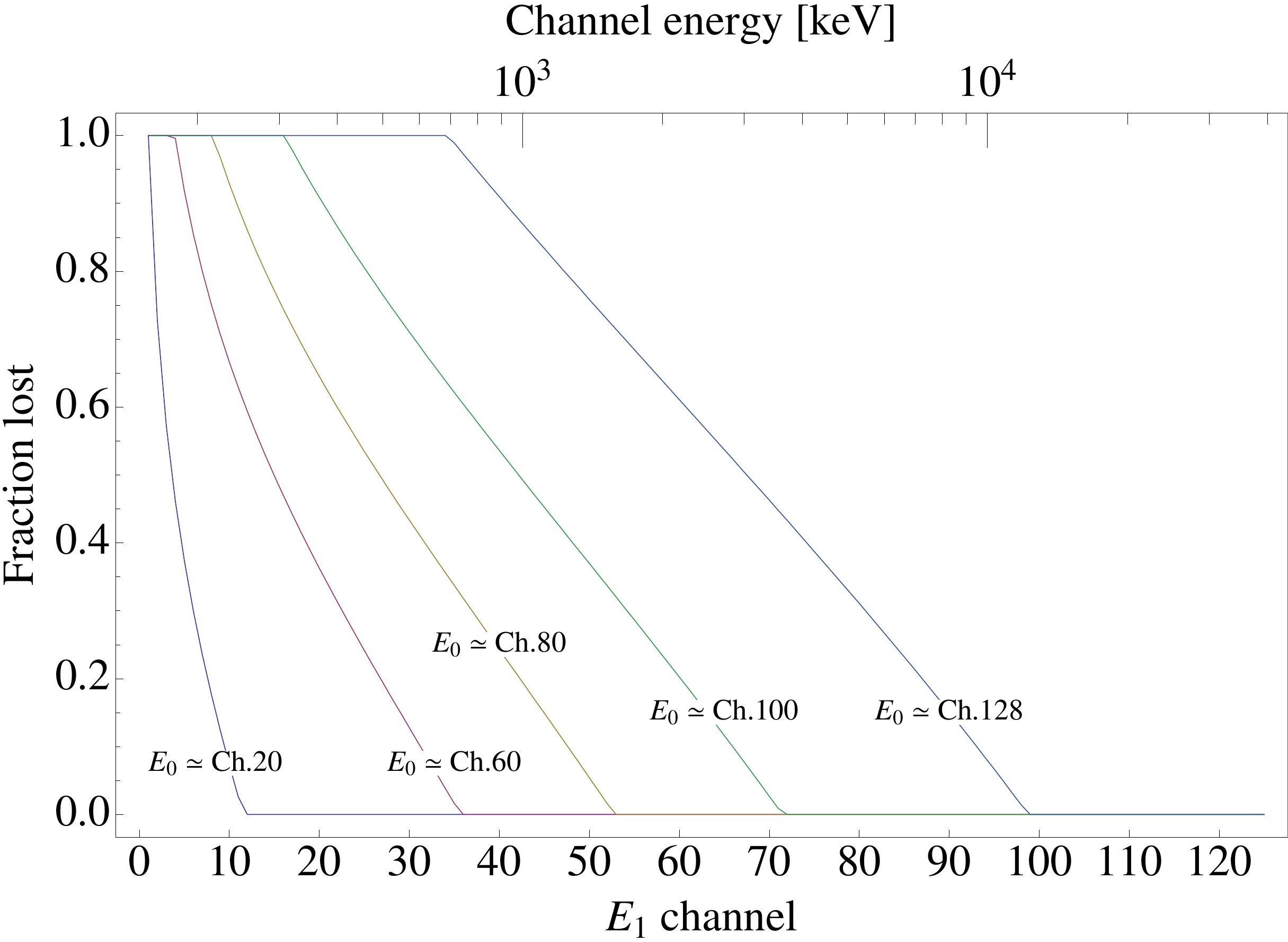}
\caption{Baseline subtraction losses due to the (first order) A+C effect.  $E_0$ is energy of the primary count.  $E_1$ is the \textit{input} energy of the second count, which occurs in the tail of $E_0$ (i.e., region C).  This is the same configuration as depicted in figures \ref{fig:ThreePanel1}(c) and  \ref{fig:ACPulseIntervals}.  As the peak amplitude $E_0$ increases, its tail becomes more negative.  If $E_1$ is too small, the summed peak in C is below registration threshold and $E_1$ is lost.  $100\%$ loss occurs when $E_0 \gg E_1$.  Each curve is calculated directly from the tail likelihood by integrating over the recorded pulse height (sum over channels in the discrete case): $Pr_{loss}(E_{0} , E_{1}) = ( 1 - \sum \limits_{\varepsilon} Pr_{A+C}( \varepsilon | E_0, E_1 ) )$}
\label{fig:TailLossCurve001}
\end{figure}

\begin{figure}[h!]
\centering
\includegraphics[width=0.95\linewidth]{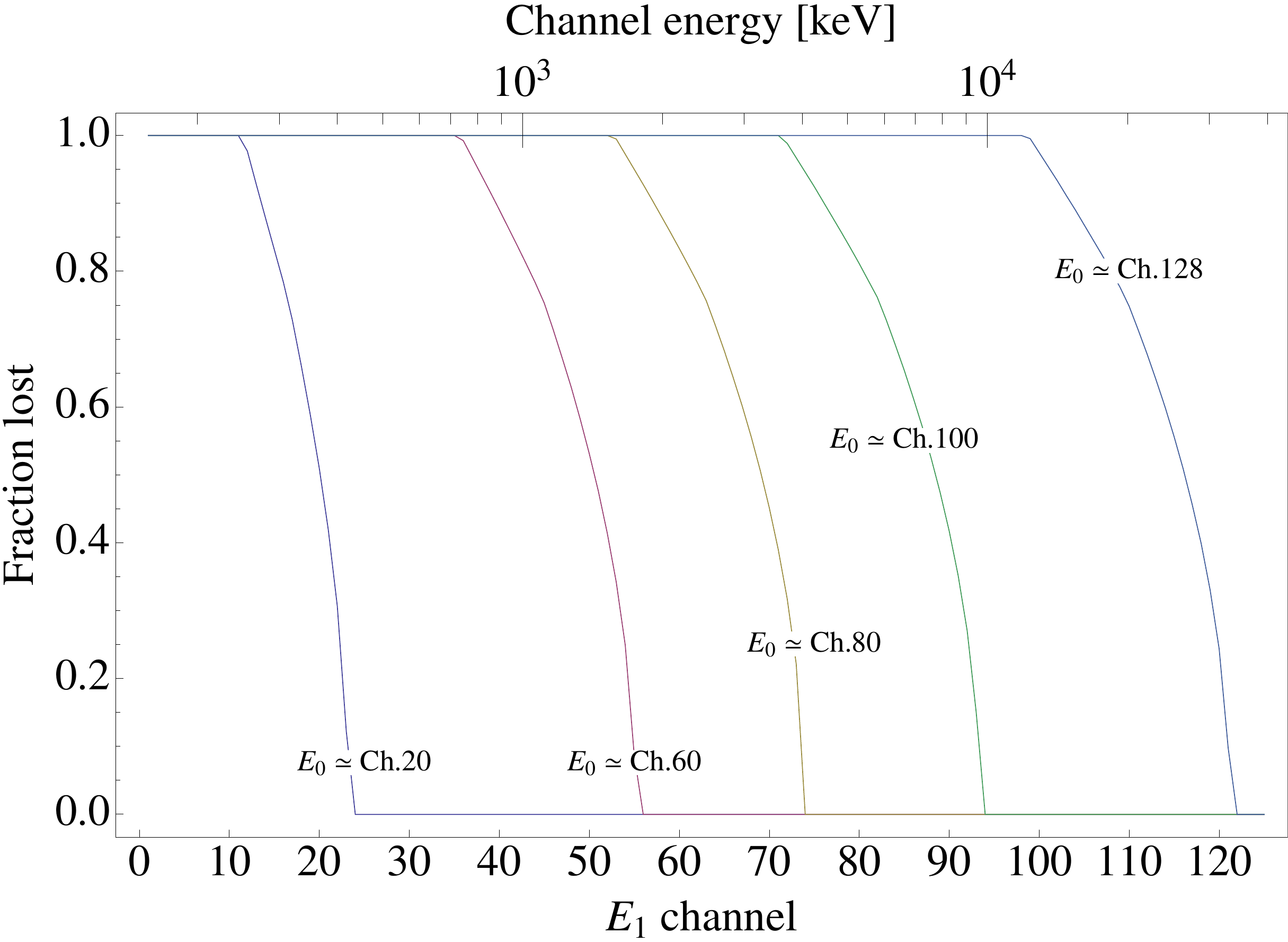}
\caption{Losses in the B+C effect (figure \ref{fig:BCPulseIntervals}).  Each curve is calculated by integrating the measurement probability in the event of $E_0, E_1$:  $Pr_{loss}(E_{0} , E_{1}) = ( 1 - \sum \limits_{\varepsilon} Pr_{B+C}( \varepsilon | E_0, E_1 ) )$. }
\label{fig:TailLossCurvex11}
\end{figure}

The second effect, tail subtraction loss, is somewhat more serious than the first in our case, because the pulse tail is longer than the peak and the negative amplitude is large.  Because of spectral dependence, tail losses can only be predicted by assuming a zeroth order pulse-height distribution $S(E)$.  These losses can be investigated by plotting marginal distributions from the tail likelihoods calculated in section \ref{TailEffect}.  Figures \ref{fig:TailLossCurve001} and \ref{fig:TailLossCurvex11} show the first-order case of the `A+C' and `B+C' pileup scenarios.  When tail spectral components are calculated using probability integrals like equation \eqref{eq:TailCorrQabc}, they turn out to have total probability less than one ($\sum_{\varepsilon} Q(\varepsilon) < 1$).  Evidently this is due to regions where tail losses exist.  By contrast the peak components are all normalized to one.

\begin{figure}[ht!]
\centering
\includegraphics[width=0.95\linewidth]{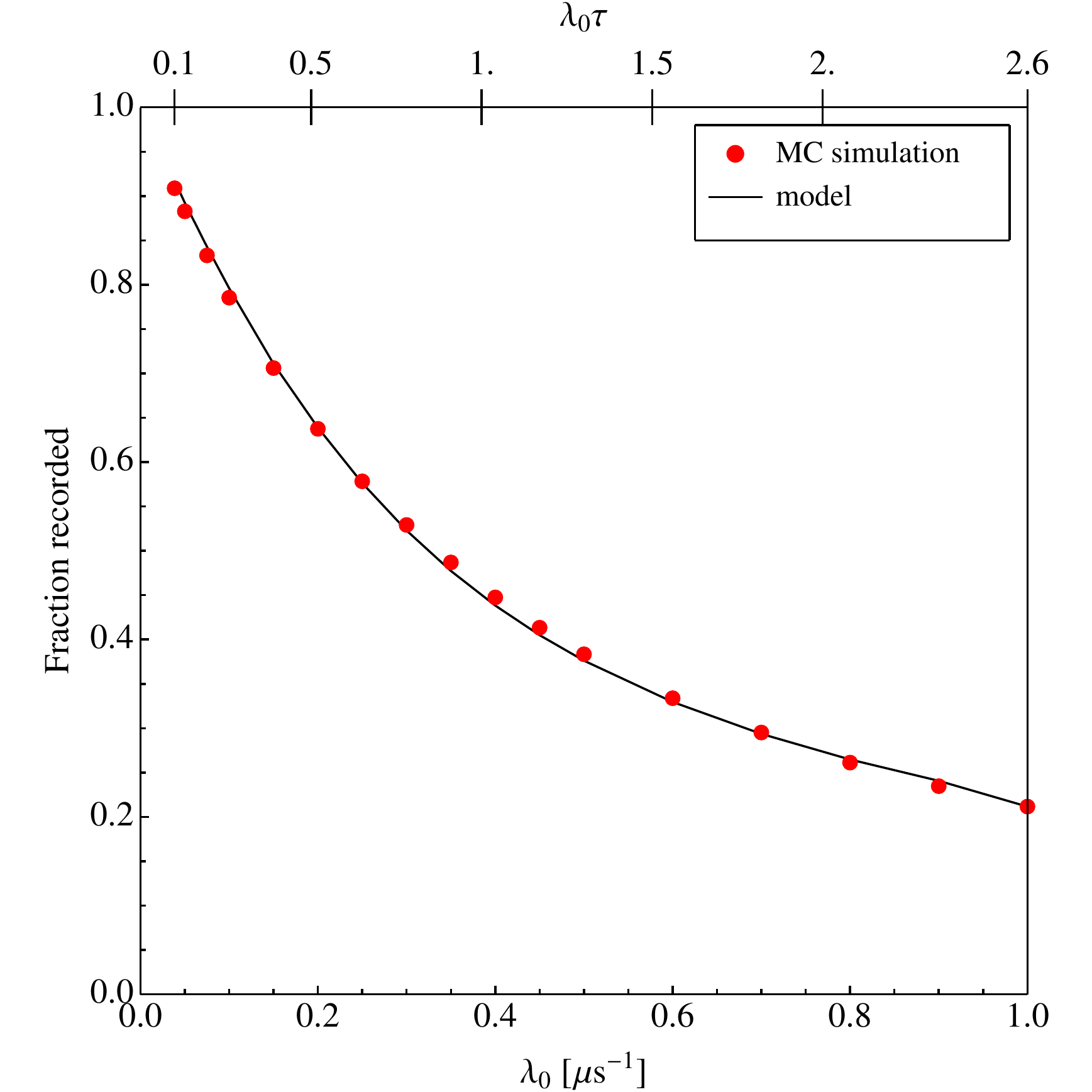}
\caption{Fraction recorded vs. detection rate.}
\label{fig:FractionVsInRate}
\end{figure}

\begin{figure}[ht!]
\centering
\includegraphics[width=0.95\linewidth]{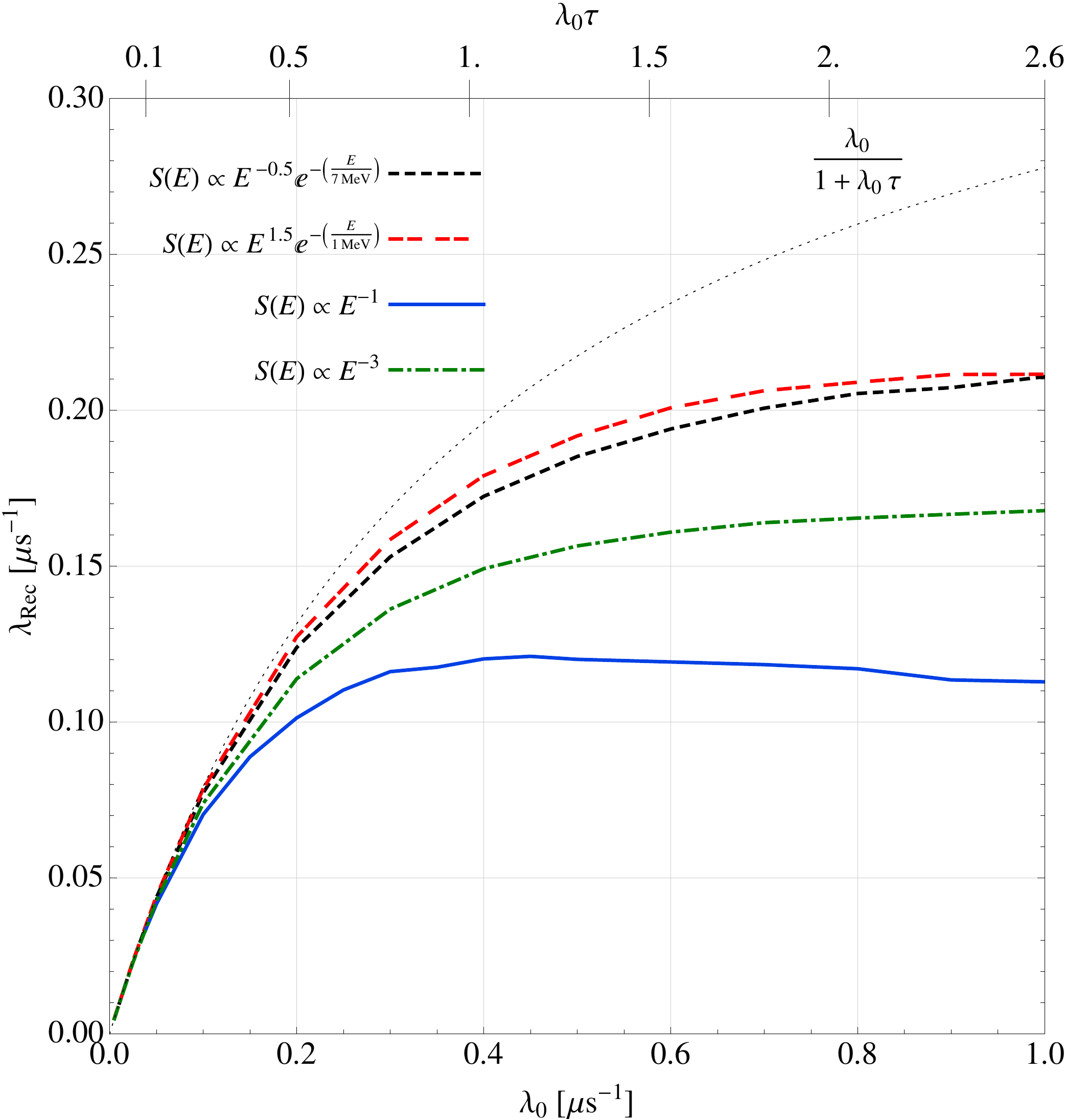}
\caption{Relation between the detection rate $\lambda_0$ and the observed rate $\lambda_{Rec}$, in millions of counts per second.  As the detection rate increases, a simple deadtime approximation becomes insufficient to explain the recorded rate.  Moreover, energy dependent losses occur due to the tail effect, which are worse for broadband, energetic (``flat")  spectra because they generate small \emph{and} large pulses.  Pulse pileup greatly increases uncertainty about the true rate given the observed rate.  As the input rate exceeds $2*10^5$ cps ($\lambda_0 \tau \approx 0.5$), constraining $\lambda_0$ given $\lambda_{Rec}$ becomes difficult (only a lower limit can be determined).  Note, since GBM detectors have their own counting electronics, $\lambda_0$ is the rate in a single detector.  Cut-off energies of 1 MeV and 7 MeV are roughly analagous to 20 keV and 150 keV in NaI detectors, respectively.}
\label{fig:OutVsInRate}
\end{figure}

The model we present gives an accurate prediction for the additional pulse-pileup losses.  Figure \ref{fig:FractionVsInRate} shows that the model result from equation \eqref{eq:FullCorrectedSpectrum} is consistent with Monte Carlo simulation.  Figure \ref{fig:OutVsInRate} shows $\lambda_{Rec} (\lambda_0 )$, demonstrating tail losses due to rate and spectral shape.

At present an analytical inversion giving $\lambda_{0} (\lambda_{Rec} )$ with deadtime and pileup has not been found, though in principle one exists until the the turnover in figure \ref{fig:OutVsInRate}.  Future efforts using a technique such as series reversion of equation \eqref{eq:FullCorrectedSpectrum} may be fruitful.  However we note that spectral fitting with the pileup correction is itself an inversion process and results in a possible solution for $\lambda_0$ given $\lambda_{Rec}$ from real data.

%
%  FIGURE 18:  
%	Fraction vs In Rate
%
%  FIGURE 19:  
%	Out Rate vs In Rate
%
%  FIGURE 20:  
%	Tail loss curve
%
%  FIGURE 21:  
%	Tail loss curve B

%%%%%%%%%%%
%
%    SECTION:
%	  
%          Discussion
%
%%%%%%%%%%%%
\section{Conclusions}
We have derived a model which accurately predicts the recorded number and spectrum of a constant intensity Poisson process with pulse-pileup, when compared to Monte Carlo simulations.  We have used the peak modeling technique and iteration method of \cite{taguchi:2010}, and extended the treatment to a three-region bipolar pulse.  The novelty in this method is that it provides a way to model input energy and timing statistics of tail pileup events.  The total spectrum is written as a state-space expansion of overlapping pulses. The technique generally applies to bipolar shaping instruments, and has been demonstrated using the true pulse shape for GBM.

\section{Acknowledgements}
This work is supported primarily by funds from the Fermi-GBM project, with additional support from the Fermi Guest Investigation (GI) program on Terrestrial Gamma-ray Flashes.  The authors would also like to acknowledge the GBM instrument team and builders for their detailed specification of the detectors and electronics.

%\end{linenumbers}

\section{References}

\bibliographystyle{model1-num-names}
\bibliography{REFERENCES}

\end{document}